\documentclass[graybox, nosecnum]{svmult}

\usepackage{mathptmx}       
\usepackage{helvet}         
\usepackage{courier}        
\usepackage{type1cm}        
\usepackage{amssymb}
\usepackage{makeidx}         
\usepackage{graphicx}        
\usepackage{multicol}        
\usepackage[bottom]{footmisc}
\usepackage{hyperref}        
\usepackage{soul}            
\hypersetup{colorlinks=true,urlcolor=blue}
\usepackage[square,numbers]{natbib}
\makeindex             

\usepackage{xspace}

\DeclareRobustCommand{\ion}[2]{%
\relax\ifmmode
\ifx\testbx\f@series
{\mathbf{#1\,\mathsc{#2}}}\else
{\mathrm{#1\,\mathsc{#2}}}\fi
\else\textup{#1\,{\mdseries\textsc{#2}}}%
\fi}

\def\arcsec{\hbox{$^{\prime\prime}$}}
\def\utw{\smash{\rlap{\lower5pt\hbox{$\sim$}}}}
\def\udtw{\smash{\rlap{\lower6pt\hbox{$\approx$}}}}

\begin{document}
\title*{Pre main sequence:  Accretion \& Outflows}
\author{P.  Christian Schneider  \thanks{corresponding author}, H. Moritz G\"unther, and Sabina Ustamujic}
\institute{P. Christian  Schneider \at Hamburg Observatory, Gojenbergsweg 11, 21029 Hamburg, Germany, \email{christian.schneider@hs.uni-hamburg.de}
\and H. Moritz G\"unther \at Kavli Institute for Astrophysics and Space Research, Massachusetts Institute of Technology,
77 Massachusetts Avenue, Cambridge, MA 02139, USA \email{hgunther@mit.edu}
\and S. Ustamujic \at INAF-Osservatorio Astronomico di Palermo, Piazza del Parlamento 1, 90134 Palermo, Italy, \email{sabina.ustamujic@inaf.it}}
%
%
\maketitle

\abstract{Low-mass pre-main sequence (PMS) stars are strong X-ray sources, because they possess hot corona like their older main-sequence counterparts. Unique to young stars, however, are X-rays from accretion and outflows, and both processes are of pivotal importance for star and planet formation. We describe how X-ray data provide important insight into the physics of accretion and outflows. First, mass accreted from a circumstellar disk onto the stellar surface reaches velocities up to a few hundred km/s, fast enough to generate soft X-rays in the post-shock region of the accretion shock. X-ray observations together with laboratory experiments and numerical simulations show that the accretion geometry is complex in young stars. Specifically, the center of the accretion column is likely surrounded by material shielding the inner flow from view but itself also hot enough to emit X-rays. Second, X-rays are observed in two locations of protostellar jets: an inner stationary emission component probably related to outflow collimation and outer components, which evolve withing years and are likely related to working surfaces where the shock travels through the jet. Jet-powered X-rays appear to trace the fastest jet component and  provide novel information on jet launching in young stars. We conclude that X-ray data will continue to be highly important for understanding star and planet formation, because they directly probe the origin of many emission features studied in other wavelength regimes. In addition, future X-ray missions will improve sensitivity and spectral resolution to probe key model parameters (e.g. velocities) in large samples of PMS stars.} 

\section{Keywords}
pre-main sequence stars, accretion, star formation, accretion shock, jet, X-ray, simulation, review

\section{Introduction}
\setcounter{footnote}{0}
How did the Sun form, how the Earth? To answer such fundamental questions, we cannot travel back in time, but we can study objects that are very similar to our Sun's progenitor: Young stellar objects that will evolve into a stellar system like ours. In this contribution, we concentrate on stellar systems broadly resembling the young Sun's, i.e., with stellar masses below about 2$\,M_\odot$.

\begin{figure}[t]
\centering
\includegraphics[width=10cm]{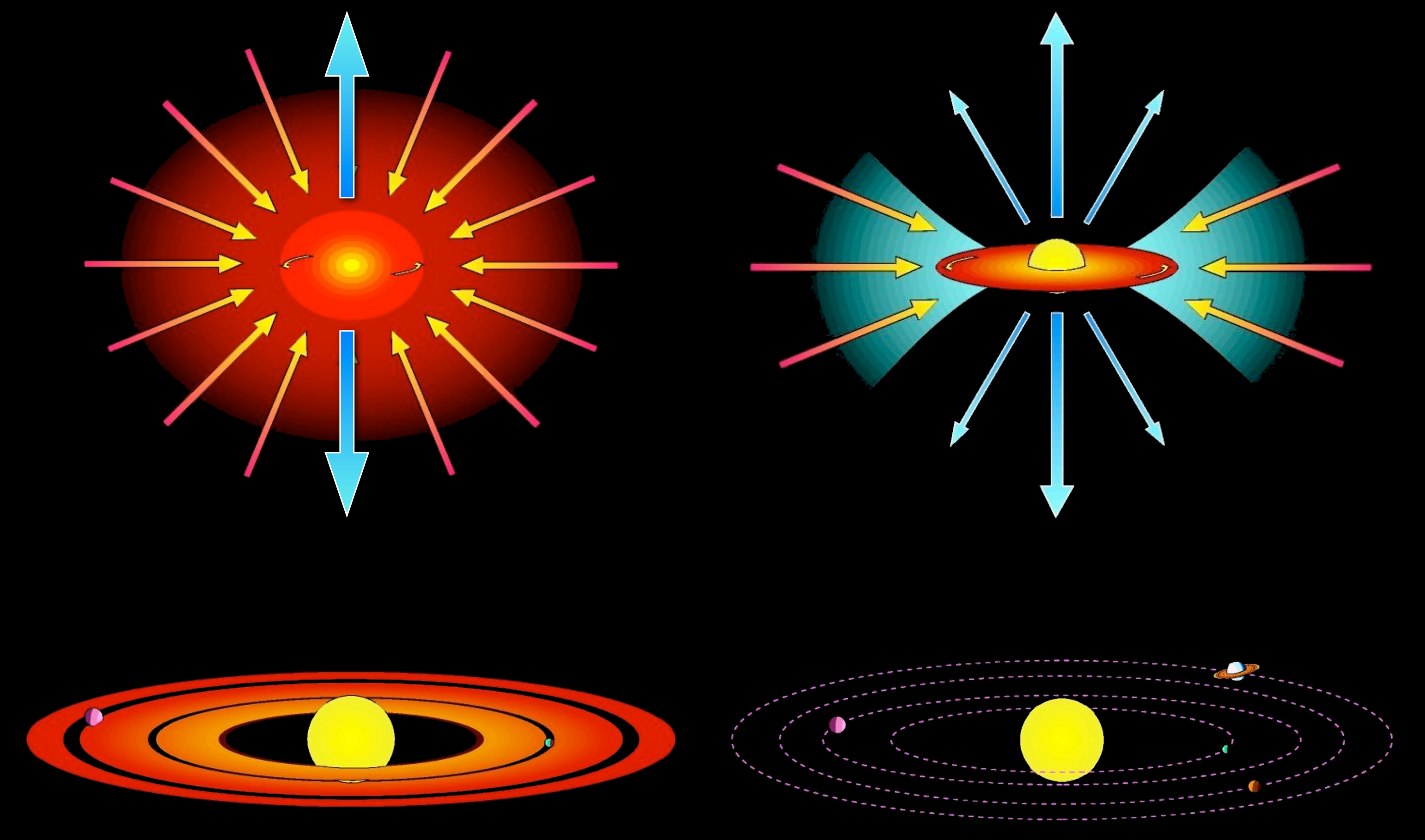}
\caption{Star formation sequence. At first, the collapse is mostly radial (top left) until the conservation of angular momentum leads to the formation of a circumstellar disk (top right). Outflows carry away angular momentum (top row). Once the envelope disperses (between the top right and bottom left stage), the mass in the disk decreases, setting the time scale for planet formation (bottom left). Eventually, only mass in the central star or the planetary system remains (bottom right). Adapted from original diagram by M.~McCaughrean. \label{fig:starform_classes}}
\end{figure}

Most, if not all, stars and planets form in molecular clouds, often associated with cool and dense filaments that collapse into protostars. In these regions, gravity dominates over the stabilizing effects of thermal pressure, turbulence, and magnetic fields \citep[e.g., ][]{McKee_2007} and protostars form  \citep{Andre_2014}. The first self-gravitating and hydrostatic core contains only a small fraction of the final stellar mass ($\sim0.01\,M_\odot$). At this time most of the mass is still in an extended envelope \citep[e.g.,][]{Gong_2015,Lee_2020}. These early objects are called class~0 \citep[see Fig.~\ref{fig:starform_classes} top left, and ][]{Andre_1993, Larson_2003}. Due to angular momentum conservation, a disk forms around the central condensation and outflows are launched, which regulate the angular momentum balance of the system. Collimated jets propagate into the interstellar medium beyond the circumstellar envelope and are often the first detectable signs of a forming star. In this phase, accretion proceeds through the disk onto the protostar while the material from the envelope replenishes the disk \citep{Padoan_2014}, which itself contains only a small fraction of the stellar mass ($\sim0.01\,M_\odot$). After roughly $10^5$\,yrs the protostar's mass rivals that of the envelope (class~{\sc i}, Fig.~\ref{fig:starform_classes} top right). The central object is hotter and more luminous than class~0 objects, and planets can form around them.

Eventually, the envelope disperses and the central star becomes visible at optical wavelengths at a stellar age of $\sim1\,$Myr; at this stage, they are called class~{\sc ii} objects or classical T Tauri stars (CTTS). The envelope dispersal mechanism is not clear yet, but winds and radiation from any nearby OB stars or otherwise radiation and outflows from low mass stars likely play a major role. Also, the winds launched by the forming stellar system itself may contribute to the envelope dispersal and reduce the envelope-to-stellar-mass conversion efficiency \citep{Frank_2014}. Accretion proceeds while planet formation continues. In this stage,  we often get a first relatively unobscured view to the newborn stars (Fig.~\ref{fig:starform_classes} bottom left). In summary, ``The fundamental problem of star formation is how stars [get] their mass'' \citep{Dunham_2014}.

Finally, the disk fully disperses and leaves behind a pre-main sequence (PMS) star, perhaps surrounded by planetary system. The central star continues to contract until it reaches the main-sequence (MS, Fig.~\ref{fig:starform_classes} bottom right). Solar mass stars reach the MS after approximately 100\,Myrs, and they reside on the MS for over 10\,Gyr.

Within this sequence of star formation, the class~II objects (CTTS) stand out, because we can see the central objects relatively unobscured and gain a detailed picture of the physical processes, including planet formation, using a large variety of observational techniques including X-ray data. However, the physical processes, in particular accretion and outflows, are thought to remain largely unchanged throughout this sequence so that we can benefit from the unobscured view during the CTTS-phase to study them.

\subsection{T Tauri Stars}
T~Tauri stars were first described as a distinct class of objects  by \citeauthor{Joy_1945} in 1945 based on their pronounced optical variability \citep{Joy_1945}. The notion that T~Tauri stars are young came with the realization that they are located to the top right of  MS stars in the Hertzsprung-Russel diagram, consistent with the expected position of stars contracting along their Hayashi tracks towards the MS \citep{Hayashi_1961}. Also, strong Li absorption lines suggest a young age, because Li is quickly depleted in stellar photospheres, i.e., high photospheric Li abundances imply that there was not enough time to process the Li accreted from the molecular cloud \citep{Magazzu_1992}.
\begin{figure}[t]
\centering
\includegraphics[width=10cm]{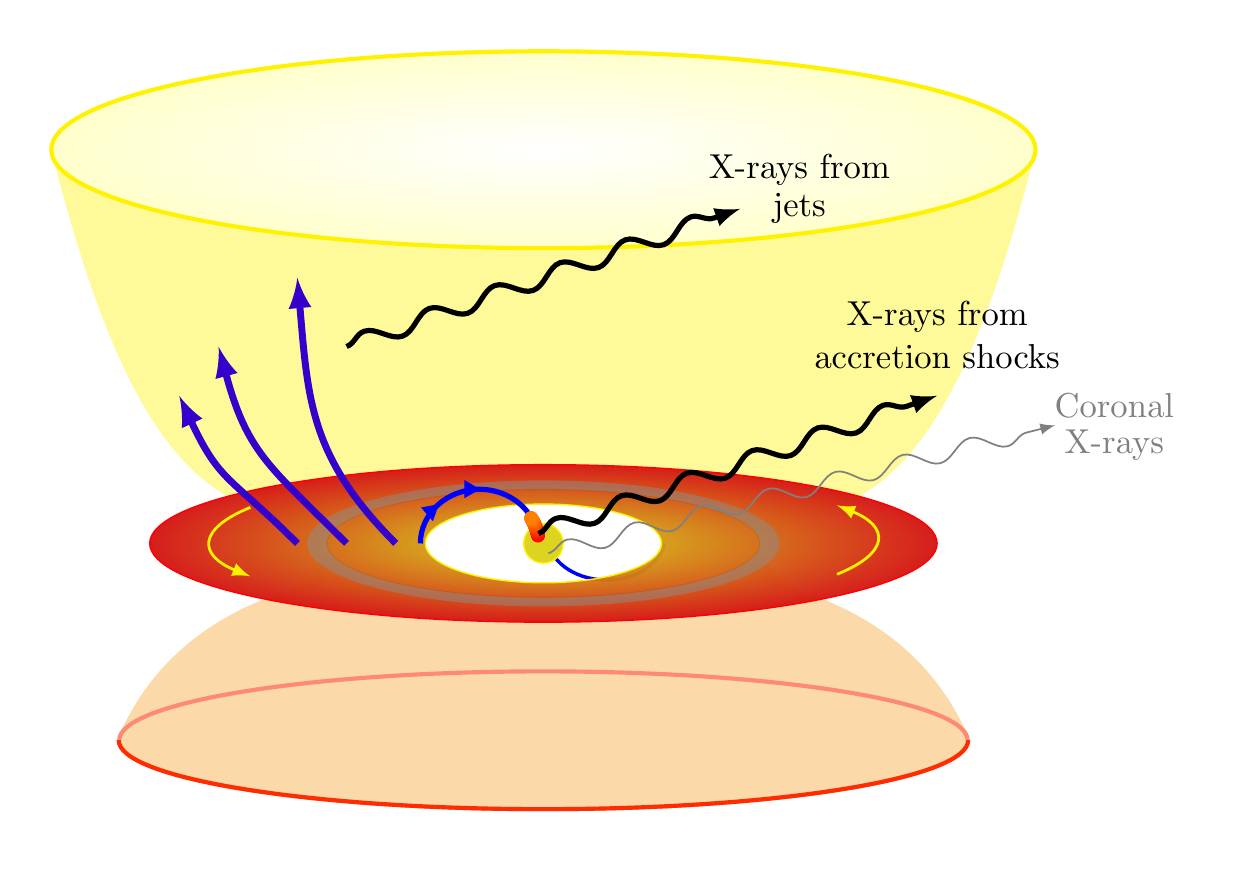}
\caption{Sketch of a classical T~Tauri system with an accretion disk and a wide-angle outflow. X-rays are coming from the stellar corona, the accretion shock and various points in the outflow. \label{fig:ctts_sketch}}
\end{figure}
On the other hand, the origin of other features like the strong emission lines such as H$\alpha$, collectively termed “chromospheric emission”, remained controversial for a long time. The discovery of the infrared excess during the 1980s and the realization that CTTS are surrounded by cold dusty disks meant that disk accretion became a leading theory explaining many features of CTTS. The  discovery that the ``chromospheric'' line profiles change from inverse to normal P~Cygni profiles and back within few days on individual systems supports this idea. It is now well accepted that accretion 
and outflows are characteristic features of CTTS \cite{Bertout_2007}, in addition to their protoplanetary disks.

Accretion and outflows are the processes that define the properties of CTTS and their disks. Figure~\ref{fig:ctts_sketch} shows a sketch of a T Tauri system with the sources of the X-ray emission discussed in this chapter marked.
During the CTTS phase, the stellar mass is already close to its final value but the interior of the star is still fully convective \citep{Stahler_2004}, which together with rather rapid rotation (few days) leads to a strong stellar dynamo and stellar magnetic fields during the CTTS phase are  in the kG range \cite{2008MNRAS.386.1234D,2010MNRAS.402.1426D}. Such strong magnetic fields disrupt the disk within a few stellar radii, because the hydrostatic star rotates slower than an approximately Keplerian disk. The interaction of the stellar magnetic field with the inner parts of the disk slows the innermost disk material down so that it cannot resist gravity and is channeled along the stellar magnetic field lines onto the star. It is accelerated to almost free-fall velocity and impacts the stellar photosphere where a strong shock forms. The details of the accretion shock are described in the next section.

The inner disk would be depleted within $10^3$ to $10^4$ years without being replenished by material from outer disk radii. Therefore, continued accretion requires radial transport of material through the disk which in turn requires the redistribution of angular momentum (AM). Several processes have been invoked to promote this AM redistribution. Prominent examples include the magneto-rotational instability \citep[MRI,][]{Balbus_1991} with non-ideal magnetohydrodynamic (MHD) extensions and disk winds \citep{Blandford_1982, Pudritz_1983}. In reality, the vertical structure of protoplanetary disks likely results in different transport mechanisms dominating at different disk heights with most of the mass being transported through the upper disk layers where magneto-centrifugally driven winds are launched.

Those winds require some net vertical magnetic field threading the disk. Material is accelerated along magnetic field lines until the magnetic field strength becomes insufficient to control the particle motion and the Lorentz force causes the collimation of the outflowing material into narrow jets. The jet velocity depends on the so-called magnetic lever arm, which relates the (approximately) Keplerian velocity at the launch radius to the jet velocity and is of order 10 \citep{Ferreira_2006}; hence, jet velocities are on the order of 300\,km\,s$^{-1}$. In addition to disk winds, there may be other outflows like stellar winds, magnetospheric ejections (from the star or the star-disk interaction region), or the so-called X-winds from the inner disk edge \citep{Shu_1994}.
In any case, it seems to be clear that magnetic fields are the main driver for outflows acceleration and collimation \cite{Fendt_2002,Fendt_2006}.

At some point, typically after a few Myrs, the disk has dispersed \citep{2009AIPC.1158....3M} and we are left with a PMS star, possibly with a number of orbiting planets. Such objects are called weak-line T~Tauri stars (WTTSs), which show no signatures of neither accretion nor outflow activity.

\subsection{The power of X-rays for studying T~Tauri stars}
The different parts of a classical T~Tauri system have grossly different temperatures: the disk midplane may be characterized by only few 10~K; the upper disk layers are likely above 1000~K; the inner edge of the disk is typically assumed to be around 2\,000~K; and stellar photospheric temperatures are about 3-4\,000~K for CTTS. Therefore, the intrinsic emission of these different components peaks in very different wavebands, which is reflected in the observational tools used to study them. Sub-mm radio to IR observations enjoy great popularity for disk studies while optical to NIR data are powerful for studying the more dynamic processes such as accretion and outflows. There are, however, some interesting exceptions, e.g., FUV data probe the hydrogen content of the inner disk as well as accretion \citep[see review in][]{Schneider_2020}.

On first sight, it may therefore appear surprising that X-ray observations play an important role for studying CTTS. However, a flow with velocities in excess of 300\,km\,s$^{-1}$, which rams into a stationary obstacle heats the post-shock plasma to temperatures in excess of 1\,MK. The radiation from such a plasma peaks in the X-ray regime and directly probes the shock region and not ``just'' reprocessed emission as seen in some other wavelengths. Thus,  X-ray studies of accreting objects are powerful diagnostics for in- and outflow processes.

\section{Accretion \label{sect:accretion}}

Accretion is the defining characteristic of star formation. It is through accretion that stars build up mass, it is through accretion that they accumulate angular momentum, and it is probably through magnetic connection between disk and accretion column that at least some part of their angular momentum is lost. Once accretions stops, stars continue to evolve, but on much longer time scales, e.g., through stellar winds, which change the mass, chemical composition, and angular momentum of a star over Gyr time-scales.

Today it is widely accepted that the accretion in T Tauri stars is magnetically funneled \citep{Hartmann_2016}. The accretion disk does not reach down to the stellar surface. Instead, it is truncated at a few stellar radii close to the radius where the disk co-rotates with the central star. While magnetic fields of young stars can be complex with the near-field dominated by higher order multipoles, the dipole moment is assumed to dominate at distances of a few stellar radii\footnote{There seems to be an evolution of magnetic field strength and geometry, where the strength of the dipole component decreases with the depth of the convective zone \cite{2012ApJ...755...97G,2019A&A...622A..72V}, but even if higher magnetic moments are stronger at the stellar surface, the dipole typically dominates at the inner disk edge.}. These (dipole) magnetic field lines couple the star with the inner disk and disk material, ionized by the UV and X-ray radiation from the star, is forced to follow the field lines. As it falls in, gravity accelerates the matter to almost free-fall velocity until it hits the surface where a strong shock develops \cite{Shu_1994}, and the post-shock plasma is so hot that it cools primarily through X-ray emission.  Depending on the height and geometry, those X-rays may or may not be visible, but the shock certainly heats the surrounding photosphere, which causes bright UV emission and optical veiling (a strong continuum that makes phtotospheric emission lines appear weaker than in a non-accreting star). To better understand the accretion geometry in CTTS, we first discuss the accretion stream and the locations of the footpoints. Then, we describe simple 1D accretion shocks before we turn to more detailed observations and models.

\subsection{The accretion stream and its foot points}
\label{sect:accretionsrteam}
The accretion stream is initially cool ($\log T\sim3-4$), because it is fed by inner disk material; it heats up as mass accelerates and comes closer to the star. The most prominent tracers are the strong and complex hydrogen emission lines. In particular H$\alpha$ is usually optically thick and often shows red-shifted absorption components compatible with near free-fall velocity \cite[e.g.][]{2000AJ....119.1881A}, and varies over time scales of hours \cite{dupree_2012}. Since many emission lines, however, do not vary with the stellar rotation period, it is likely that the inner disk is more important for accretion than the anchor point on the stellar surface \cite{2021A&A...649A..68S}. 

The structure of the stellar magnetic field can be probed by Zeeman-Doppler imaging and, using certain assumptions, those fields can be extrapolated out to the inner disk edge where the dipole component usually dominates the coupling of the star with the accretion disk; the accretion funnel therefore follows the dipolar field lines. On the stellar surface itself, Doppler imaging can locate the position of the accretion funnels which are often found near the pole, e.g.\ in BP Tau \cite{2008MNRAS.386.1234D} or V2129 Oph \cite{2011A&A...530A...1A}, but sometimes at lower latitudes as in V2247 Oph \cite{2010MNRAS.402.1426D}. 
Simulations can reproduce the analytical model of accretion foot points near the pole in the dipolar field, but they also point to more complex geometries when disk, stellar rotation, and stellar magnetic field are not aligned \cite{2021MNRAS.506..372R}.

Accretion is a rather dynamic process. As the star and the disk rotate, and the magnetic field and the disk structure evolve, the accretion geometry and the accretion rate can change on time scales as short as minutes or as long as centuries; accretion can also switch off temporarily or permanently, as the star looses its disk. Nevertheless, the basics of the  accretion process remain similar as long as the star accretes from the disk, and we expect that X-rays are produced by the same processes throughout; perhaps with the exception of FU~Ori type outburst, which may involve major reconfigurations of the disk \cite{2014prpl.conf..387A}.

\subsection{X-ray signatures of the accretion shock}
\label{sect:accretionobs}
Any accretion generated X-ray emission in young stars must be seperated from the ubiquitous X-ray emission by coronal activity in low-mass stars. Young stars rotate faster and thus have more pronounced magnetic activity and stronger coronal X-rays than older stars such as our Sun. Therefore, accretion signatures are challenging to distinguish from coronal emission in broad-band X-ray spectra with their limited spectral resolving power (e.g., CCD-type X-ray spectra). To the contrary, one rather finds an inverse correlation between the integrated X-ray flux and the accretion rate \cite{2005ApJS..160..401P, Schneider_2018}. However, this does not contradict the idea that accretion shocks generate soft X-rays as models show that the shock would mostly contribute below 1~keV \cite{1999AstL...25..430L} and those soft X-rays are easily absorbed by circumstellar material or the remnants of the star forming cloud.

Therefore, additional X-ray diagnostics are needed to test the existence of accretion generated X-ray emission and the following four findings strongly suggest that accretion-powered X-rays are seen in CTTS. 

\begin{figure}[t]
\centering
\includegraphics[width=0.49\textwidth]{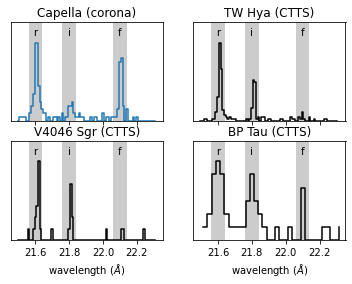}
\includegraphics[width=0.49\textwidth]{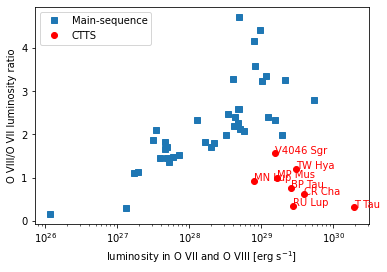}
\caption{Signatures of accretion in classical T Tauri stars from high-resolution grating spectroscopy. \emph{Left:} Density-sensitive O~{\sc vii} triplet. Capella is a MS star with an $f/i\sim 4$, while the other three sources are examples of CTTS with $f/i < 1$. All lines are unresolved; they appear wider in BP~Tau because the data is taken with a lower-resolution spectrograph (\emph{Chandra}/LETG, while the other three are \emph{Chandra}/HETG). \emph{Right:} Ratio of O~{\sc viii} to O~{\sc vii} line flux compared to the total flux in oxygen lines. All accreting sources are well offset from the MS stars, indicating additional soft emission plasma. Modified from Ref.~ \cite{2013ApJ...771...70G} (see there for data sources). \label{fig:softexcess}}
\end{figure}

First, high-resolution grating spectroscopy allows one to measure the density of the emitting plasma from line ratios in the O~{\sc vii} and Ne~{\sc ix} triplets\footnote{These triplets resemble the electron configuration of Helium and are also called He-like triplets.}. \emph{XMM-Newton} and \emph{Chandra} carry X-ray gratings with sufficient spectral resolution to separate the three lines of the density sensitive triplets: A resonance line ($r$), an intercombination line ($i$), and a forbidden line ($f$). In a collisionally excited plasma, the $f$ line is typically stronger than the $i$ line, but collisions in a high-density plasma or strong UV fields can excite an electron from the upper level of the $f$ line to the upper level of the $i$ line. However, strong UV fields are only relevant for A or B stars and not for lower-mass classical T~Tauri stars discussed here. For CTTS, a low $f/i$ ratio is thus a sign for high densities in the emission region. Because coronal emission typically comes from low density plasma \citep[$n_e\lesssim10^{10}$\,cm$^{-3}$, e.g.,][]{Ness_2002}, X-ray emission from a high density plasma likely originates behind the shock front of an accretion shock, where densities are thought to be much higher than in the corona ($\gtrsim10^{13}$\,cm$^{-3}$). Figure~\ref{fig:softexcess} (left panel) shows examples for three CTTS which all show $f/i < 1$ compared to a typical MS star with $f/i\sim 4$. A triplet indicative of high densities was first seen in TW~Hya \cite{Kastner_2002}, and the same pattern has since been confirmed in a number of CTTS. Low $f/i$-ratios are therefore strong evidence for accretion generated X-rays.

Second, a lower O~{\sc viii}/O~{\sc vii}-ratio is another feature that sets CTTS  apart from WTTS, their non-accreting sibblings (see Fig.~\ref{fig:softexcess} right). This indicates that the X-ray emission in CTTS is produced by an, on average, cooler plasma compared to non-accreting stars of comparable luminosity\footnote{Non-accreting stars show a correlation between plasma temperature and X-ray luminosity so that one must correct for this to find any accretion-driven effect in plasma temperature.} \cite{2007A&A...473..229R,2007A&A...474L..25G}. Another view of this finding is that 
CTTS have an intrinsic (coronal) O~{\sc viii}/O~{\sc vii}-ratio in line with low-luminosity WTTS or MS stars since fainter coronae are also cooler. This view implies that CTTS have additional cool plasma compared to normal coronae, and this additional plasma may be accretion-powered as the post-shock plasma radiates more strongly in O~{\sc vii} than in O~{\sc viii} at the expected temperatures.

A third observation is that CTTS usually show abundances that are Ne enhanced and Fe depleted compared to solar abundances \citep{Stelzer_2004}. This pattern is also seen in active stars, where it is attributed to element separation in the corona due to different values of the first ionization potential (FIP) of ions. Neon has a particularly high FIP and iron a particularly low one, so the pattern observed in active stars is called IFIP (inverse FIP). However, CTTS are cooler than MS stars of comparable luminosity and so alternative scenarios have been discussed \citep{Drake_2005}. Since accretion comes from the inner edge of the accretion disk, it is possible that disk processes separate elements. If grain-forming elements condense into grains, pebbles, and proto-planets, the material at the inner disk edge might be depleted of Fe, Si, and similar metals, while noble gases stay in the gas phase and are thus preferentially accreted. 

Fourth, X-ray grating spectra of TW Hya have sufficient S/N to determine  line shifts for some strong emission lines. The lines from cooler plasma show a $38.3 \pm 5.1$~km/s shift compared with the hotter, coronal lines \cite{2017A&A...607A..14A}. And while this indicates different origins for the cool and hot plasma, it is also much less than the free-fall velocity. This implies that we must see the shock almost perpendicular to the line-of-sight. TW~Hya is observed close to pole-on, so the accretion shock(s) seen in X-rays must the located in the equatorial region for TW~Hya.

\subsection{Physics of accretion in 1D}
\label{sect:accretionphysics}
With X-ray observations pointing to a different emission pattern in accreting stars compared to non-accreting ones, we discuss simple models to check if these X-ray signatures are broadly compatible with X-ray emission from the accretion shock.

Many aspects of the accretion physics can be described in a 1D model where all mass motion is parallel to the magnetic field lines. We review some of the basic physics of accretion columns and accretion shocks. While modern models go far beyond such a simple prescription, the foundation of all accretion shock models still is to convert the gravitational energy in the disk to kinetic energy of the free-falling gas, which in turn is used to heat the plasma in the accretion shock and powers the radiative cooling.

The free fall velocity $v_{\textnormal{free}}$ of material coming from an inner disk radius of $R_\mathrm{in}$ onto a star with mass $M_*$ and radius $R_*$ is
\begin{equation}
v_{\textnormal{free}} = \sqrt{{2GM_*} \left(\frac{1}{R_*} - \frac{1}{R_\mathrm{in}}\right)} \approx 620 \sqrt{\frac{M_*}{M_\odot}}\sqrt{\frac{R_\odot}{R_*}} \frac{\textnormal{km}}{\textnormal{s}}\ \label{eqn:freefall}
\end{equation}
where $G$ is the gravitational constant. The inner radius is typically a few stellar radii and thus $v_\textnormal{free}$ is very close to infall from infinity.

\subsubsection{The shock front}
We start our discussion with stationary shocks and ignore all turbulent fluxes. In the shock front, ions and electrons are heated differently, but they remain strongly coupled and reach the same temperatures within a few mean-free path lengths---a region so thin that it is justified to treat them as a single fluid.

Somewhere along the accretion column, a shock forms when the forward ram pressure becomes comparable to the pressure of the underlying material. The shock front itself is very thin, only of the order of a few mean free paths \cite{raizerzeldovich}. Therefore it can be treated as a mathematical discontinuity described by the Rankine-Hugoniot jump-conditions \cite[][chap.~7, \S~15]{raizerzeldovich}; in the shock the super-sonic infall velocity is converted mostly into thermal energy. Since we assume the direction of flow parallel to the magnetic field, the Lorentz force does not influence the dynamics. Marking the state in front of the shock front by the index 0, and that behind the shock by index 1, the Rankine-Hugoniot conditions become
\begin{eqnarray}
\rho_0 v_0 &=& \rho_1 v_1 \label{RH1}\\
P_0+\rho_0 v_0^2 &=& P_1+\rho_1 v_1^2 \label{RH2}\\
\frac{5 P_0}{2\rho_0}+\frac{v_0^2}{2}&=&\frac{5 P_1}{2\rho_1}+\frac{v_1^2}{2} \ ,\label{RH3}
\end{eqnarray}
where $v$ is the velocity, $\rho$ the total mass density of the gas and $P$ its pressure.

From the jump conditions, the shocks will heat gas to a temperature $T$
\begin{equation}
kT \simeq \frac{3}{16}\mu m_p v^{2} \approx 0.3\,{\rm keV}\left(\frac{v}{500\,{\rm km/s}}\right)^{2} \approx3.5\times10^6\,{\rm K} \left(\frac{v}{500\,{\rm km/s}}\right)^{2},
\label{eqn:Tshock}
\end{equation}
where $\mu$ is the dimensionless atomic weight.

\subsubsection{Structure of the post-shock region}

In the post-shock region the gas emits radiation and cools down, so the energy of the gas is no longer conserved.  However, the particle number flux $j$ of ions (and atoms)
\begin{equation}j=nv\label{j_n}\end{equation}
is conserved, where $n$ is the ion/atom number density; the electron number density is denoted by $n_{\mathrm{e}}$. The total momentum flux $j_p$ is conserved, since we ignore the momentum loss by radiation:
\begin{eqnarray}
j_p&=&\mu m_{\mathrm{H}} n v^2+P \nonumber \\
   &=&\mu m_{\mathrm{H}} n v^2+nkT \label{j_p}
\end{eqnarray}
where $m_{\mathrm{H}}$ denotes the mass of a hydrogen atom.

Let us next consider the energy balance in the post-shock region. In general,
\begin{equation} \label{tsminuspdvisdu} T d\Sigma -P dV=dU \end{equation}
where $\Sigma$ denotes the entropy and $U$ the internal energy of the plasma. The quantity $T d\Sigma=dQ$ denotes the heat flux through the boundaries of the system. Here, the energy loss $Q_{col}$ proceeds through collisionally excited higher electronic states, which decay radiatively.

Assuming that the shock location is stationary, we get $\frac{d}{dt}=\frac{\partial}{\partial t}+\frac{\partial z}{\partial t}\frac{\partial}{\partial z}=v\frac{\partial}{\partial z}$ depending on the location $z$, measured from the shock front inwards; differentiation with respect to $z$ will be indicated by $'$.
The internal energy $U$ is in this case the thermal energy $U=\frac{3}{2}kT$, the pressure $P$ can be rewritten using the equation of state. The specific volume $V$ is the inverse of the number density $V=\frac{1}{n}$.
It is convenient to write the electron number density as \mbox{$n_{\mathrm{e}}=x_e n$,} with $x_e$ denoting the number of electrons per heavy particle so that Eq.~\ref{tsminuspdvisdu} becomes
\begin{equation}
\label{energyelec}
v\left(\frac{3}{2}x_e k T_{\mathrm{e}}\right)'+v x_e n k T_{\mathrm{e}} \left(\frac{1}{n}\right)'=-Q_{col} x_e n\,.
\end{equation}

\begin{figure}[t]
\centering
\includegraphics[width=\textwidth]{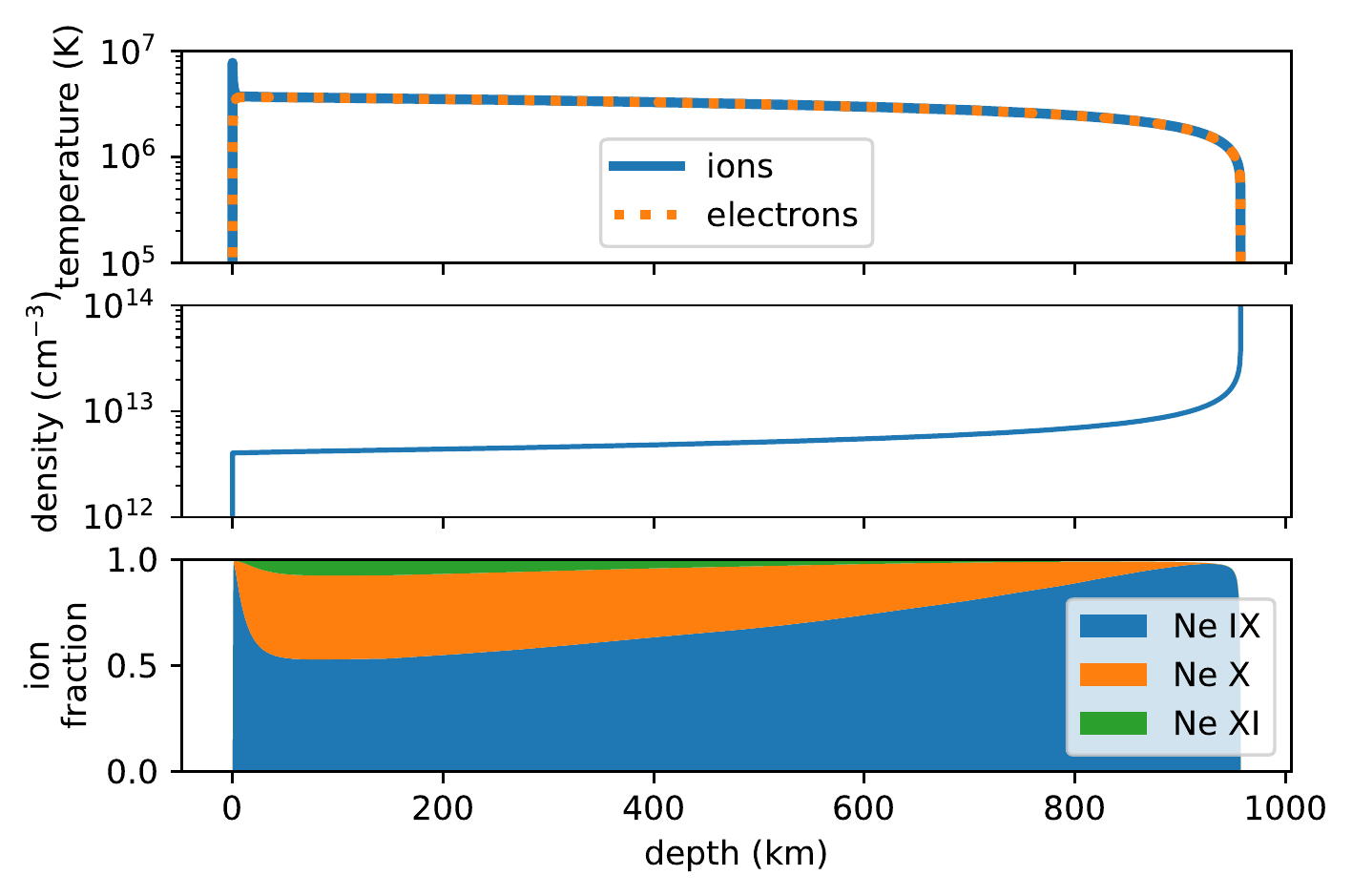}
\caption{Structure of the accretion shock in a 1D model. Pre-shock velocity is $v_0=525$~km~s$^{-1}$ and density $n_0=10^{12}$~cm$^{-3}$, typical for T Tauri stars. Ions and electrons reach temperature equilibrium fast, ionization equilibrium takes longer. Data from Ref.~\cite{Guenther_2007}. \label{fig:1dshock}}
\end{figure}

We now have $n$, $v$, and $T$ as variables and three hydrodynamic (HD) equations (\ref{j_n}, \ref{j_p}, and \ref{energyelec}), so the structure of the post-shock region can be calculated (see figure~\ref{fig:1dshock} for an example).

Simulations based on these or very similar equations show that the accretion shock produces plasma matching the observed X-ray temperatures \citep{lamzin_1998} and the total energy in the accretion stream can be determined from fitting UV and optical spectra to determine the mass accretion rate \citep{calvet_1998}. Observations of the density-sensitive line ratios in He-like triplets such as in Fig~\ref{fig:softexcess} (left) can be explained by 1D models assuming a combination of accretion shock and coronal plasma \cite{Guenther_2007} and at the same time give a mass accretion rate that can be compared to observations of UV and optical tracers.

\subsection{Why we need to go beyond 1D models}

One dimensional models are successful in many aspects, but there are also observational and theoretical arguments suggesting that important physical processes are not captured in 1D. In deep \emph{Chandra} observations of TW~Hya, densities can be measured from three different triplets covering a range of peak formation temperatures. The density is highest in Mg~{\sc xi} and lowest in O~{\sc vii}, although Mg~{\sc xi} is formed at a higher temperature \cite{Brickhouse_2010}. Because the post-shock region is isobaric, one would expect Mg~{\sc xi} emission from a region directly behind the accretion shock and O~{\sc vii} from denser layers deeper down ($P\propto T$), i.e., opposite to what is observed. One possible explanation for this conundrum is that the observed O~{\sc vii} emission is not from the accretion shock itself, but originates in hot material that escapes the accretion column to the side and is denser than a normal stellar corona, but not as dense as plasma immediately behind the accretion shock. That in turn means that the O~{\sc vii} flux may not be a good measure of the total accretion rate even when corrected for coronal contributions.

This is corroborated by the surprising observation that X-ray determined mass accretion rates, often from O~{\sc vii} data, are very similar for most sources despite a difference in optically determined mass accretion rate by three orders of magnitude \cite{2011A&A...526A.104C}. The inconsistency between estimates based on data from different wavelengths can be explained if the accretion streams are not homogeneous structures but have a density profile, and the inner layers either form shocks deeper in the atmosphere or simply have their X-ray emission reprocessed by the outer layers of the accretion stream \cite{Sacco_2010,Schneider_2018,2021Natur.597...41E}.

Time variability can give us further insight. In V4046~Sgr the emission lines from soft, presumably shock-heated, plasma have a period of exactly half the orbital period of the close binary. Together with Doppler-imaging this leads to the interpretation that we can observe the shock best when the accretion flow is perpendicular to the line of sight, while the accretion funnel blocks the view of the shocked region at other times \cite{2012ApJ...752..100A}. 
There are several classes of variable PMS accretors, including FU~Ors and EX~Ors, where the accretion rate changes by orders of magnitude or stars where a change in the disk structure moves absorbing material into our line of sight, such as in RW~Aur or also in AA~Tau. However, those changes happen on much larger scales than the accretion shock itself and are not discussed here any further.

Finally, observations of our Sun provide the opportunity to look for situations analogous to accretion onto young stars, although the corresponding accretion rate is obviously much lower. The benefit of such data is that spatially resolved data in the UV and EUV, even if not in X-rays, are available with long time coverage and high cadence while most observations of the accretion shock on young stars are spatially unresolved. One particularly interesting observation is the event that happened on June, 7$^{th}$, 2011 when parts of an erupting filament fell back into the Sun \cite{2013Sci...341..251R} (see also \cite{Reale_2014}). The infall speed of up to 450~km/s was comparable to free-fall accretion onto T Tauri stars and the initial infall triggered upflows that shocked with later fragments causing UV emission on the Sun; such a process would not be covered in a 1D approach.

\begin{figure}[t]
    \centering
    \includegraphics[width=11cm]{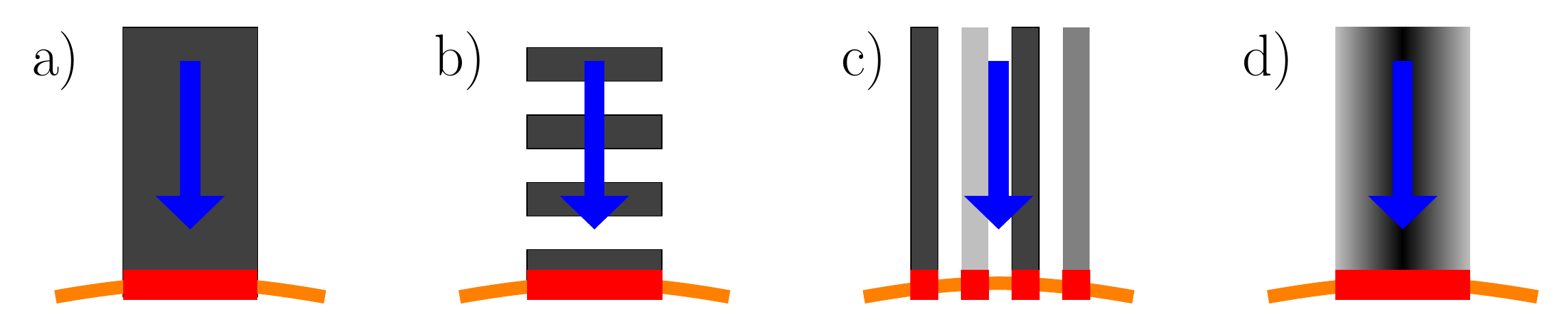}
    \caption{Different scenarios proposed for the 
    structure of accretion columns. The~stellar surface is at the bottom (orange) and 
    material flows along the magnetic field lines onto the stellar surface (in the direction indicated 
    by the blue arrow). The~accretion shock (red) forms on the stellar surface. 
    Density is shown in gray scale. From left to right: 
    a) One homogeneous column with one density and a single infall velocity; 
    b) one ``column'' is decomposed into individual blobs that are, individually, homogeneous in density;
    c) multiple columns that are individually homogeneous, i.e.,~have different densities but are otherwise equal; 
    d) one column with a density stratification (high density in the center of the column indicated by the black region, lower density in the outer region indicated by the grayish color) and one infall velocity. 
    }
    \label{fig:column}
\end{figure}

A correct description of the coupling between radiation and hydrodynamics is necessary to account for the effects of absorption and emission of radiation, because photons and mass may travel differently through the domain. Indeed, improved models including local thermodynamic equilibrium (LTE) radiation transfer now show that the heated photosphere does not radiate as a simple black body in the optical and infrared \cite{Dodin_2012,Dodin_2013}, but also produces line emission, which selectively fills in some photospheric absorption lines, possibly biasing accretion rate measurements based on optical veiling.
Similar optical depth effects also
have a non-negligible result on the typical characteristics of the 
accretion dynamics, on the estimation of its X-ray surface luminosity 
\cite{Sa_2019}, on the heating of the accretion column \cite{1999AstL...25..430L,Costa_2017}, and the predicted lightcurve \cite{2021ApJ...908...16R}. Thus, we now turn to the 3D structure of the accretion process.

\subsection{The multi-D structure of the accretion shock}

\begin{figure}[t]
    \centering
    \includegraphics[width=9cm]{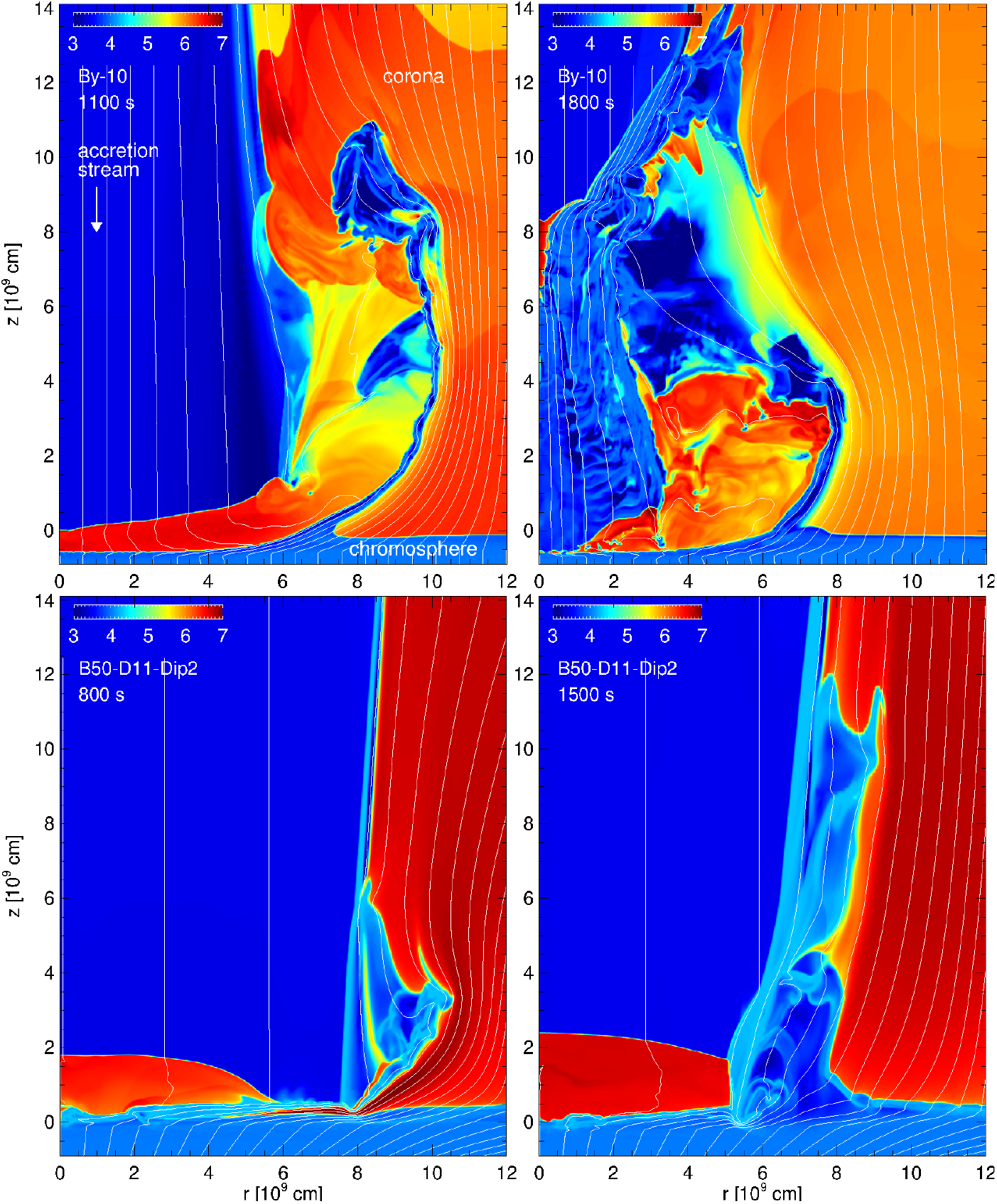}
    \caption{Temperature distributions (in log scale) of the accretion stream at the labeled times (increasing from left to right) for two different models: simulation By-10 from \cite{Orlando_2010} (upper panels), and simulation B50-D11-Dip2 from \cite{Orlando_2013} (lower panels). The initial position of the transition region between the chromosphere and the corona is at $z = 0$. The white lines mark magnetic field lines. Figure courtesy of S. Orlando.}
    \label{fig:accretion}
\end{figure}

The 3D density structure of the accretion columns is key to correctly describe accretion and 
different scenarios have been proposed (see Fig.~\ref{fig:column}). The first models considered one homogeneous column with one density and a single infall velocity (Fig.~\ref{fig:column}\,a). 
\citet{Orlando_2010, Orlando_2013} modelled a constant and uniform accretion stream that propagates along the magnetic field lines considering uniform  (see upper panels in Fig.~\ref{fig:accretion}) and nonuniform  (see lower panels in Fig.~\ref{fig:accretion}) stellar magnetic fields at the impact region of the accretion columns where a hot slab  with temperatures up to $\approx 5\times10^6$~K forms (see Fig.~\ref{fig:accretion}). After impact, part of the shock column is buried under a column of optically thick material and may suffer significant absorption \cite{Orlando_2010,Orlando_2013}.
The structure, dynamics and stability of the accretion shock strongly depend on the configuration and strength of the magnetic field \cite{Orlando_2010}, and thus on the plasma $\beta$ defined as $\beta =$~gas pressure/magnetic pressure. 
The post-shock region is efficiently confined by the magnetic field for shocks with $\beta \lesssim 1$ , while strong outflows of shock-heated material 
escape laterally for shocks with $\beta$ above $10$ \cite{Orlando_2010,Orlando_2013}. For intermediate cases (see left upper panel in Fig.~\ref{fig:accretion}), the escaping material is kept close to the accretion column 
if the magnetic field is strong enough and gradually surrounds the primary stream until it eventually perturbs the accretion column (see right upper panel in Fig.~\ref{fig:accretion} and \cite{Orlando_2010}). 
A nonuniform magnetic field causes a large field component perpendicular to the stream velocity, and the resulting additional magnetic pressure at the base of the stream limits the plunging of the slab into the chromosphere (see left lower panel in Fig.~\ref{fig:accretion}) and rather pushes chromospheric material sideways along the magnetic field lines to the upper atmosphere (see right lower panel in Fig.~\ref{fig:accretion} and \cite{Orlando_2013}). 

\begin{figure}[t]
    \centering
    \includegraphics[width=12cm]{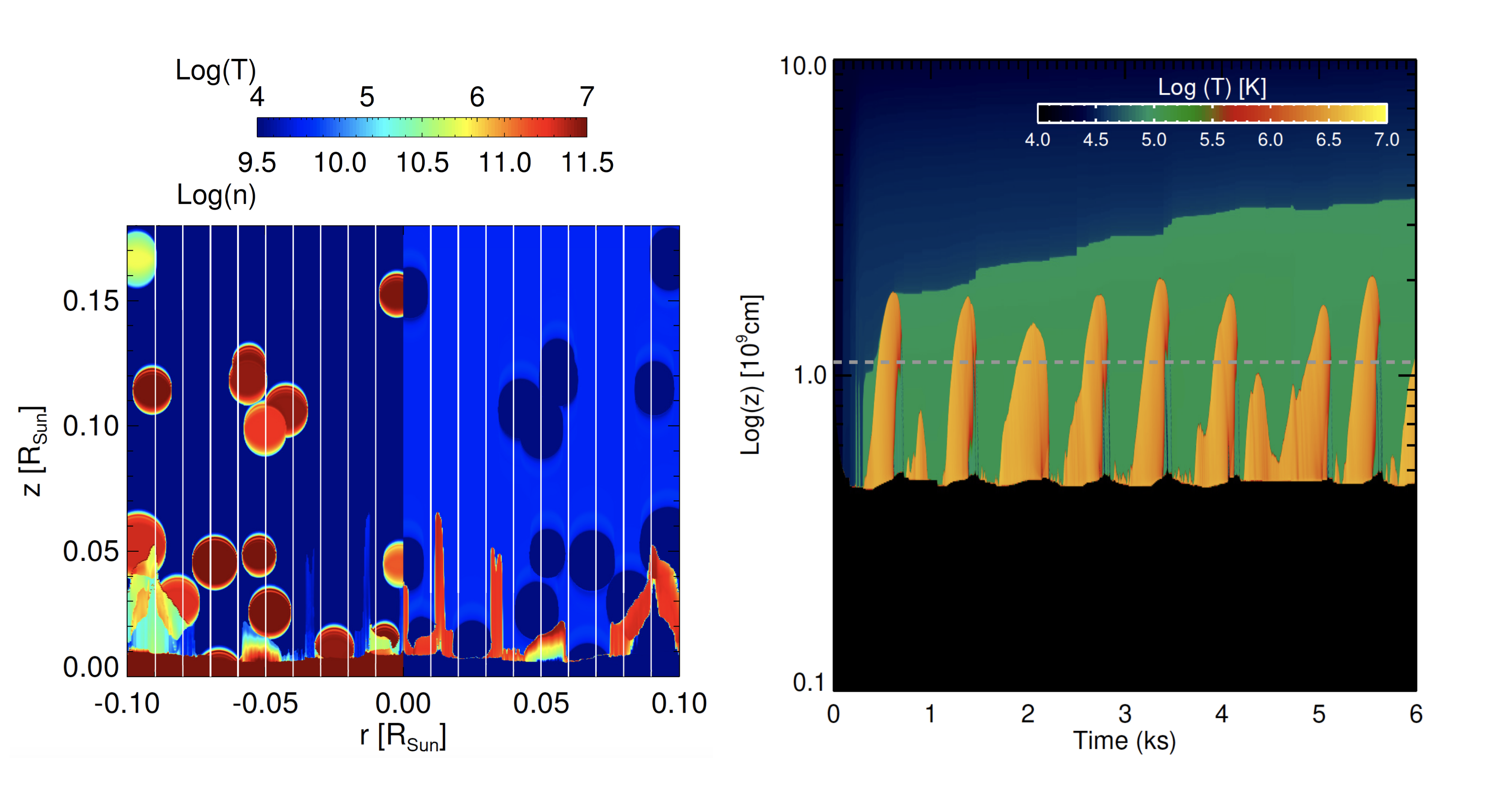}
    \caption{\textbf{Left}: Color maps of evolution of density (left half-panel) and temperature (right half-panel) of plasma for the general case of fragmented stream explored in \cite{Colombo_2016}. White lines represent magnetic field lines. Figure credit: \citet{Colombo_2016}, A\&A, 594, 93, 2016, reproduced with permission © ESO. 
    \textbf{Right}: Space-time map, in logarithmic scale, of temperature for the non-LTE model run by \cite{Colombo_2019b}. The green region corresponds to the hottest part of the radiative precursor of the shock. The gray
    dotted line marks the pre-impact position of the 
    chromosphere. Figure credit: \citet{Colombo_2019b}, A\&A, 629, 9, 2019, reproduced with permission © ESO.}
    \label{fig:colombo}
\end{figure}

The temperature resolved density structure \cite{Brickhouse_2010}, high filling factors of the cool plasma \cite{2007A&A...468..529G,Schneider_2018}, and  fragmented material observed in solar flares  \cite[see][]{2013Sci...341..251R}, all point towards more complex accretion streams.
Examples are density stratification and clumpy structures formed by several blobs (Fig.~\ref{fig:column} c \& d and, e.g., \cite{Matsakos_2013,Colombo_2016}). Perturbations in the accretion column (namely a clumpy stream and a oblique impact) can disrupt the shock structure and create a more inhomogeneous post-shock region with the post-shock region still strongly dependent on the plasma $\beta$ \cite{Matsakos_2013}.
Randomly fragmented accretion streams (see left panel in Fig.~\ref{fig:colombo}) produce reverse shocks that propagate upflows through the unshocked fragmented stream when the blobs impact onto the stellar chromosphere \cite{Colombo_2016}.
As a result, the structure of the post-shock region is very complex and consists of several knots and filaments of plasma with a wide range of velocities, densities, and temperatures \cite{Colombo_2016}.

Both models and observations indicate that optically thin and thick plasma coexist in the impact region of accretion streams so that radiative transport phenomena are important for the observed emission.
For instance, an envelope of dense and cold material may develop around the shocked column, influencing the observability of the shock-heated plasma in the X-ray band (\citep{Orlando_2013}, see also Fig.~\ref{fig:accretion}). The X-ray emission from the shocked region can be heavily reduced due to local absorption by material around the hot plasma, e.g.,  by the unperturbed stream or the perturbed chromosphere.
The magnitude of this
effect depends sensitively on the viewing angle as this changes the amount of material along the line of sight towards the hot plasma, and may skew density estimates derived from He-like triplets, as shown by simulations considering absorption effect when calculating the synthetic X-ray emission  \cite{Bonito_2014}.

Absorption of the shock emission may also (pre-) heat part of the material in the accretion region that has not gone through a shock yet. This effect can alter the emission escaping from the system including the X-ray diagnostics. Correctly describing the complex interplay between dynamics and radiation, however, requires MHD simulations coupled with radiative transport in the non-LTE regime. Performing such calculations, \citet{Colombo_2019b} showed  that about 70\,\% of the radiation emerging from the hot slab can be absorbed by the pre-shock material in the funnel, which gradually heats up forming a radiative precursor of the shock in the accretion column (see green region in right panel in Fig.~\ref{fig:colombo}). 

In addition to MHD models, laboratory experiments are now available, which can be scaled to accretion shocks on CTTS since the MHD equations are invariant under certain scaling transformations. Nowadays, laboratory experiments employ magnetic fields, densities, temperatures, and other hydrodynamical parameters that can be scaled to the stellar case. 
One example of such an experiment consisted of a collimated narrow laser-produced plasma stream that propagates along a strong external magnetic field, very similar to accretion on CTTS \cite{Revet_2017}. Similarly to the MHD simulations described above, this laboratory experiment shows
that, upon impact, the plasma is ejected laterally from the accretion shock and then refocused by the magnetic field towards the incoming stream, forming a shell that envelops the shocked core \cite{Revet_2017}. This hot shell provides an additional absorber that reduces the observable  X-ray emission. Also, experiments with accretion streams impacting a tilted surface have been performed, which correspond to accretion streams along complex magnetic field structures. The resulting plasma flow is highly asymmetric and a large amount of plasma escapes laterally from the accretion flow, showing poor confinement of the accreted material and reduced heating compared to the normal
incidence case \cite{Burdonov_2020}.

\subsection{Variability and accretion outbursts}
In the previous sections, we concentrated on CTTS with their well-ordered magnetically-funneled accretion streams. There are, however, other classes of PMS accretors, which very rare and therefore more challenging to study. One such class are objects, which show accretion outbursts with the accretion rate suddenly jumping by several orders of magnitude and then decaying back over time scales from months (EX~Or objects) to decades or centuries (FU~Or objects) \cite{2014prpl.conf..387A}. EX~Or outbursts are not only shorter, they can also be observed to be repetitive. These eruptive stars show bright and hot coronal emission, possibly because the coronal structures are compressed with a higher density or because having a bright corona in the first place somehow triggers the accretion burst \cite{2011ApJ...741...83T,2019ApJ...883..117K}. Most sources are highly absorbed, but soft emission compatible with an accretion shock has also been observed \cite{2010A&A...522A..56G}---again much brighter than in CTTS due to the increased accretion rate. 
On the other hand, in CTTS with their much lower accretion rates, we have no evidence of any correlation between magnetic flares and accretion events \cite{1997A&A...324..155G,2019ApJ...876..121E}.

\subsection{Towards a coherent picture of the accretion shock}
X-ray observations of CTTS show features unique to accreting stars, primarily a cool ($\sim10^6\,$K), high-density ($n_e\gtrsim10^{12}\,$cm$^{-3}$) plasma excess. This excess emission is compatible with expectations for a high-velocity shock, i.e., emission from the post-shock region of the accretion shock. However, uniform, homogeneous accretion columns produce a post-shock structure incompatible with the X-ray data: the measured densities significantly diverge from predictions for the post-shock region \cite{Brickhouse_2010}, the mass-accretion rates derived from X-ray grating spectra are too low \cite{2011A&A...526A.104C}, and the covering fractions on the stellar surface are too large \cite{Schneider_2018}. Therefore, more complex accretion funnels are needed to reconcile observations with models. The leading idea is currently a density stratified or fragmented accretion stream \cite{Colombo_2016}, where the strength and configuration of the magnetic field plays a fundamental role. Also, absorption by optically thick material and the effect of radiative heating of pre-shock material brings models closer to observations \cite{Colombo_2019b}. In summary, a dynamic and structured accretion funnel is the most likely explanation for the observed emission, because this naturally leads to multiple regions with different properties (e.g., density and temperature) that contribute to the (X-ray) emission; and those different emission regions are necessary to reproduce observations.

\section{X-rays from protostellar jets}

\begin{figure}[t]
 \centering
\includegraphics[width=6cm]{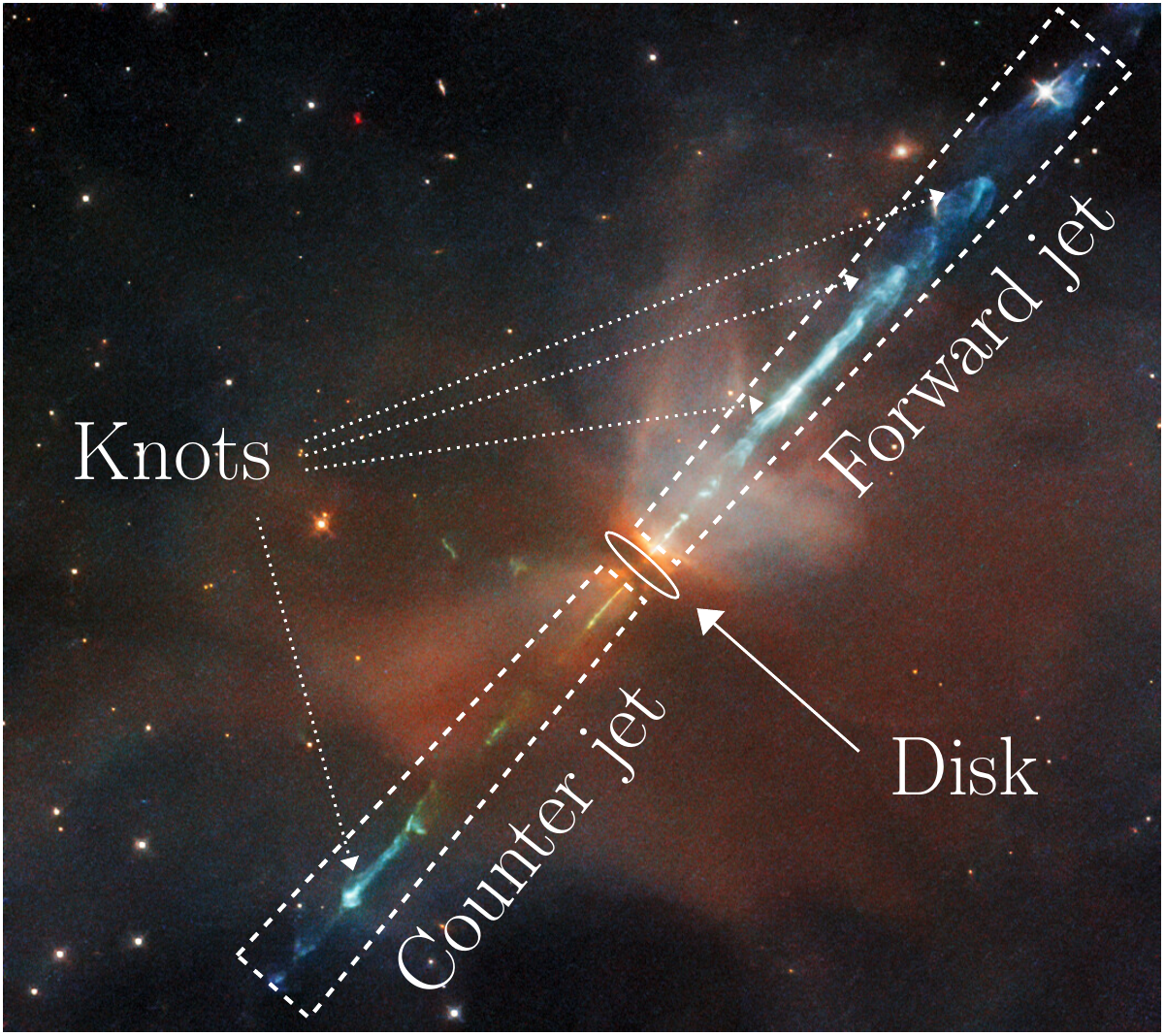}
\caption{HST composite image of HH~111 with some relevant features labeled. Backround image credit: ESA/Hubble \& NASA, B. Nisini. \label{fig:HH111} }
\end{figure}

Protostellar outflows  were ``discovered''\footnote{The nebulosity around T~Tau was first described in the 19th century  by \citet{Burnham_1890, Burnham_1894}.} by \citet{Herbig_1950,Herbig_1951} and \citet{Haro_1952,Haro_1953} in the early 1950s. These authors studied the optical, nebular emission around young stars. This ``nebular'' emission is typically concentrated in distinct emission regions along a relatively straight path extending away from a young star, called knots, and mostly studied in forbidden emission lines (FELs, see Fig.~\ref{fig:HH111}). These emission knots are called Herbig-Haro (HH) objects and trace plasma shock-heated by supersonic material ejected from a young stellar system.
Individual emission knots move away from the driving source with typical velocities of a few hundred km\,s$^{-1}$ \cite{Eisloffel_1994}.

Outflows in PMS stars are detected in a wide range of wavelengths and show very different morphologies, from jets with opening angles of only a few degree to less-collimated, wide-angle outflows. 
They are seen as long as the central star is accreting, and the ratio between outflow and accretion rates of roughly $\dot{M}_{out}/\dot{M}_{acc}\sim0.1$ appears constant throughout this phase \citep[e.g.,][]{Cabrit_2007, Nisini_2018}. This implies that the mass-loss rate through the outflow decreases with time, because the accretion rate decreases with time and the most powerful jets are launched early during the star formation process. Nevertheless, the current picture is that jet physics remain the same during the entire accretion phase.

Protostellar outflows are powered by a variety of mass ejection phenomena occurring during the star formation process, which are intimately related to the accretion process. 
Many prominent outflow phenomena are compatible with magneto-centrifugally launched disk winds \cite{Blandford_1982,Pudritz_1983} from the inner regions of protoplanetary disks \citep[roughly 0.1\,au to 10\,au, see][]{Frank_2014}. Disk winds have initially large opening angles and flow velocities are moderate (roughly a few tens of km\,s$^{-1}$).  The outflowing disk material is then accelerated along magnetic field lines to velocities of a few hundred km\,s$^{-1}$ and the Lorentz force causes the collimation of the outflow to narrow opening angles within the innermost tens of au from the driving source. Disk winds naturally explain that the highest flow velocities are observed close to the jet axis while subsequently lower velocities are found for the outer streamlines \cite{Bacciotti_2000}.

There are other possible launch regions in addition to disk winds, too. First, stellar winds can launch from the hot stellar corona much like the solar wind \citep{Matt_2005}. Thus, such stellar winds are very hot ($\gtrsim10^6$\,K) and may cool via X-ray emission. Second, magnetospheric ejections can launch from the region between the star and the inner edge of the disk \citep{Zanni_2013}. Third, X-winds could be launched from a narrow region close to the inner edge of the disk \citep{Shu_1994}. Even more mechanisms to launch outflows may exist  during the earliest stages of star formation, e.g., in the the class~0 phase, but they produce slower outflows and are therefore not related to the observed X-ray emission and ignored here.

The physical properties of protostellar jets such as density and ionization can be derived from specific line ratios between FELs  \cite{Bacciotti_1999}. From these FELs, we know that shocks within the outflow heat the plasma to temperatures of roughly $10^4\,$K. According to Eq.~\ref{eqn:Tshock}, such a post-shock temperature corresponds to shock velocities of $\sim20-30$\,km\,s$^{-1}$, which is only a fraction of the flow velocity (a few hundred\,km\,s$^{-1}$).
Because of the low densities, $\lesssim10^4\,$cm$^{-3}$ beyond few tens of~au from the driving source, radiative cooling of the shock heated plasma is dominated by hydrogen (e.g., H$\alpha$) and metallic FELs such as [O~{\sc i}], [S~{\sc ii}], and [Fe~{\sc ii}].
Nevertheless, protostellar jets have been studied at UV over optical to (near-) IR and radio wavelengths, where even non-thermal emission has been seen \cite{Ainsworth_2014,Rodriguez_2017}, though $\gamma$-rays from protostellar jets have not been detected yet.
Still, there were no indications for plasma temperatures of $T\gg10^5\,$K when the first detections of protostellar jets in X-rays were made so that these discoveries came to a surprise.

\subsection{X-ray observations of jets}

\begin{figure}[t]
\centering
\includegraphics[width=11cm]{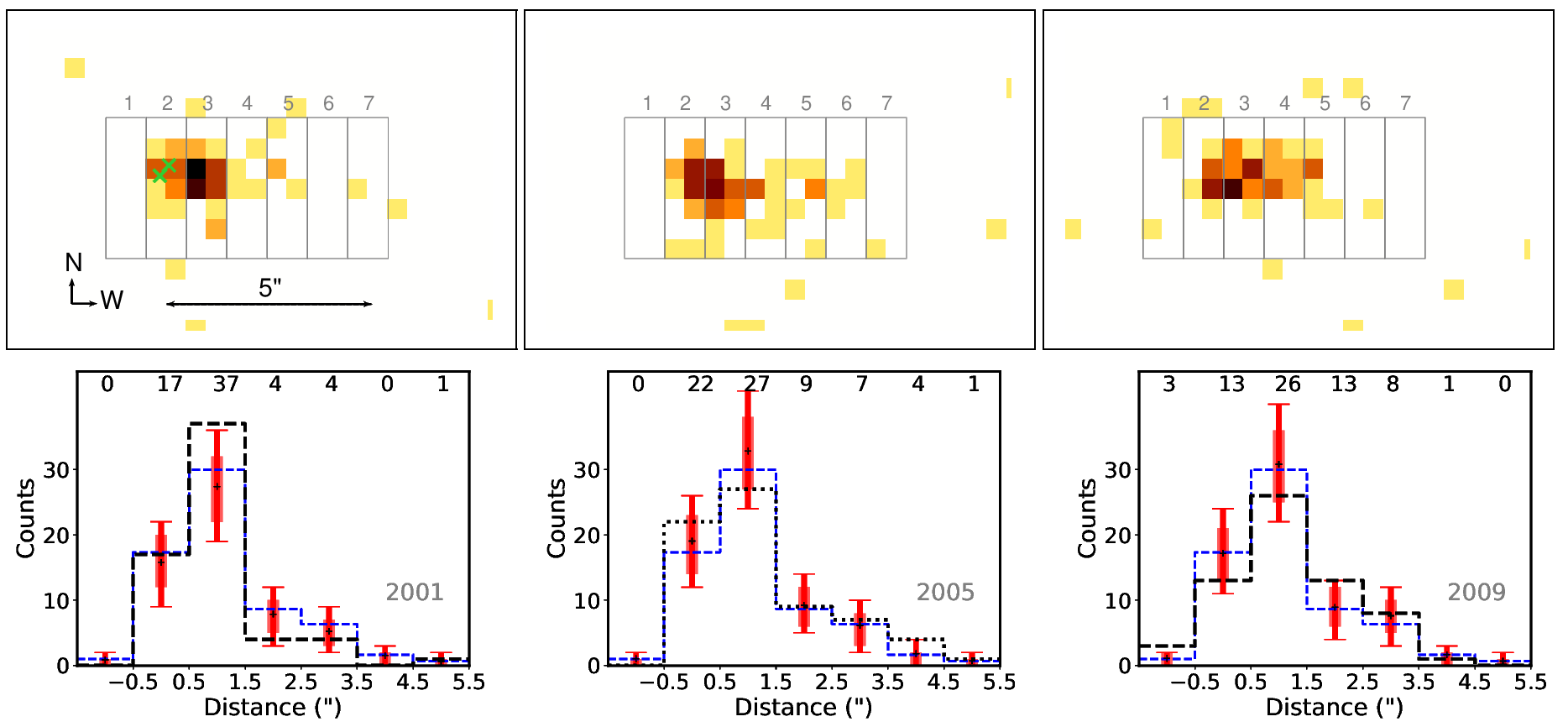}
\caption{X-ray emission from HH 154 observed with \emph{Chandra} at three different epochs. {\bf Top}: X-ray images within the 0.5 to 3.0\,keV photon energy range. The driving source(s) are located in stripe 2 and the approximate positions of the two binary components are marked by the green crosses. {\bf Bottom}: Count number histograms along the jet axis, i.e., counting photons within the stripe indicated in the top three panels. The solid curve describes the mean of the three observations; the blue dotted line indicates the mean ignoring efficiency differences between the three observations. The red bars indicate the expectation
value for a time constant emission, shown at 70\% and 90\% probability ranges. Adapted from \citet{Schneider_2011}, A\&A, 530, 123, 2011, reproduced with permission © ESO. \label{fig:hh154}}
\end{figure}

The first X-ray detections of protostellar jets were obtained almost simultaneously in 2001 by \citet{Pravdo_2001} using \emph{Chandra} (HH~2) and \citet{Favata_2002}
using \emph{XMM-Newton} (HH~154 or L\,1551~IRS\,5, see Fig.~\ref{fig:hh154}); both outflows are driven by class~I objects. The association of the \emph{Chandra} detected X-ray emission to HH~2 is straight forward, because the X-ray emission  (a) spatially coincides with one of the optically brightest knots, (b) appears extended, and (c) the X-ray spectrum is soft ($T\sim1.2\times10^6\,$K) unlike typical coronae of young stars (or brackground objects). In contrast, the association of the X-rays from the L\,1551~IRS\,5 complex to HH~154 strongly builds on the severe extinction towards the protostellar sources and the relatively low extinction towards the jet \citep[$A_V(jet)\sim10\,$mag vs $A_V(protostars)\gtrsim150$\,mag, e.g.,][]{White_2000,Fridlund_2005}. This high $A_V$ towards the stellar sources points to a jet origin of the X-rays despite the proximity of the X-ray emission to the stellar position and the comparably high mean X-ray energies ($\sim1\,$keV), simply because any reasonable X-ray emission from the stellar system would not make it directly to the observer though the thick absorber. Later, \citet{Bally_2003}, located the X-rays from HH~154 to the base of the outflow using new \emph{Chandra} data, compatible with a location where the outflow is collimated to the narrow jet observed further away from the sources. 

These two early discoveries of X-ray emission  from protostellar jets already indicate two different phenomena, namely emission from the jet base (HH~154) and from working surfaces at large distances to the driving sources, i.e., from individual jet knots (HH~2). To simplify the discussion, we separate these two phenomena here.

\subsubsection{X-rays from jet knots}

We start our discussion with the X-ray emission from jet knots at distances well beyond the jet collimation region and the protostellar envelope ($d\gg100\,$au, i.e., $\gg1$ arcsec for the nearest jets), because this allows us to derive the physical requirements to produce significant X-ray emission within outflows. Examples of this sub-class of X-ray emitting jets are HH~2 \citep{Pravdo_2001}, HH~80/81 \citep{Pravdo_2004}, TKH~8 \citep{Tsujimoto_2004}, the resolved  jet from HD~163296 \citep[HH~409, see ][]{Swartz_2005,Guenther_2013},   HH~168 \citep{Pravdo_2005,Schneider_2009}, HH~210 \citep{Grosso_2006}, HH~216 \citep{Linsky_2007}, Z~CMa \citep{Stelzer_2009}, and possibly, HH~248 \citep{Lopez_2015}; the driving sources of these jets range from low to high mass objects suggesting that 
the source details do not dominate the jet's appearance in X-rays. However, the relatively high fraction of high-mass objects driving X-ray emitting jets may indicate that their jets more frequently favor X-ray emission.

The jet's X-ray emission is typically seen close to emission at other wavelengths from cooler plasma   \citep[$T\sim10^4$\,K, e.g.,][]{Pravdo_2004,Grosso_2006,Schneider_2012}.  In some cases, the X-ray emitting knots have been reported to also show [O~{\sc iii}] emission \citep[$T\sim10^5$\,K, see ][]{Grosso_2006} and the X-ray emitting knots tend to be among the fastest knots with space velocities of $\gtrsim400$\,km\,s$^{-1}$ \citep[e.g.,][]{Pravdo_2001}. 

Multi-epoch \emph{Chandra} observations show that the X-ray emission in HH~2 moves along the jet path with high velocity \citep{Schneider_2012} similar to the optical knots.  Therefore, shocks are thought to be the most likely explanation for the X-ray emission  although the X-rays may not come from the leading working surface in all cases, but might be displaced somewhat towards the driving sources \citep[][]{Pravdo_2005}. Shock heating requires some residual velocity of the knots, but proper motion measurements are challenging in X-rays, because of the diffuse nature of the emission, the relatively low X-ray count numbers, and the relatively large distances\footnote{Many targets are farther than 500\,pc, so that displacements on the sky are small.}. Cooling or fading of X-ray emitting knots may have been seen in the jet of HD~163296 \cite{Guenther_2013}, but the count number is small and multi-epoch data are mostly consistent with a constant X-ray flux.

A key requirement to produce significant X-ray emission from shocks is a sufficiently high flow velocity so that high-velocity shocks are possible and the post-shock plasma is hot enough to emit significantly in the X-ray range. The derived temperatures for the X-ray emitting plasma are roughly  in the low to medium MK range so that shock velocities $v_{shock}$ of $500\pm200$\,km\,s$^{-1}$ are required (the physics of shock heating is the same as in the accretion shock, see ~Eq.~\ref{eqn:Tshock}). Such shock velocities may result where the jet flow encounters low-velocity  material in front of it, either from previous ejecta or dense clumps in the molecular cloud. This is compatible with the proper motion measurement for the X-ray emitting knot in HH~2: the interaction of a fast (490\,km\,s$^{-1}$) knot overtaking a slower one (220\,km\,s$^{-1}$) results in shock and post-shock velocities compatible with the measured values \cite{Schneider_2012}.

The second requirement is a sufficiently high X-ray luminosity. The observed X-ray luminosities range from a few $10^{29}$\,erg\,s$^{-1}$ (HH~2) up to $5\times10^{31}$\,erg\,s$^{-1}$ (HH~80). We follow \citet{Raga_2002} to estimate if such X-ray luminosities are plausible for protostellar jets. The radiative loss of an optically thin, homogeneous plasma is 
\begin{equation}
L = EM\, \Lambda(T) \approx n_e^2\, V\,\Lambda(T)  \label{eqn:L}
\end{equation} 
with the emission measure $EM$, the radiative loss function $\Lambda$ and with $n_e$, $V$ having the same meaning as in the previous section (electron density and emitting volume $V$), and we have assumed $n_H\approx n_e \approx n$ for our order of magnitude estimate. We assume that the emitting volume can be characterized as a cylinder of radius $r_s$ and a height given by the cooling length of the shocked plasma $d_{cool}$, i.e., the  emitting volume is $V = \pi r_s^2 d_{cool}$. This is obviously a simplified picture, because the post-shock region is neither isothermal nor is the density constant; nevertheless this approach is sufficient to illustrate the general picture here. A reasonable estimate for $d_{cool}$ is the interpolation formula provided by \citet{Heathcote_1998} for the \citet{Hartigan_1987} shock models
\begin{equation}
d_{cool} = 2.24\times10^7 \textrm{cm} \left(\frac{v_{shock} }{\textrm{km\,s}^{-1}} \right)^{4.5} \left(\frac{\textrm{cm}^{-3}}{n_0} \right)\,, \label{eq:dcool}
\end{equation}
with the pre-shock density $n_0$ so that equation~\ref{eqn:L} becomes
\begin{eqnarray}
L & = & \pi r_s^2  \, d_{cool} \, n_e^2\,\Lambda_X(T) \label{eq:L}\\
  & = &  1.1\times10^{37} \textrm{cm}^3 \left(\frac{r_s}{10^{14}\,\textrm{cm}}\right)^2 \left(\frac{v_{shock} }{\textrm{km\,s}^{-1}} \right)^{4.5} \left(\frac{n_0}{\textrm{cm}^{-3}} \right) \, \Lambda(T)
\end{eqnarray}
where we have used $n_e\approx4n_0$ appropriate for strong shocks. A typical value for $\Lambda(T)$ in the 1--10\,MK range is a few $10^{-23}$\,erg\,s$^{-1}\,$cm$^{-3}$ so that we may expect an X-ray luminosity of
\begin{equation}
L \sim 1\times10^{29} \left(\frac{r_s}{100\,\textrm{au}}\right)^2 \left(\frac{v_{shock}}{300\,\textrm{km}\,\textrm{s}^{-1}}\right)^{4.5} \left(\frac{n_0}{100\,\textrm{cm}^{-3}} \right) \, \left( \frac{\Lambda(T)}{3\times10^{-23} \textrm{erg}\, \textrm{s}^{-1} \textrm{cm}^{-3}} \right)\,,
\end{equation}
which is within a factor of two of the \citet{Raga_2002} estimate for radiative shocks. Therefore, realistic shocks within protostellar jets can produce detectable X-ray emission, and faster jets produce much more X-ray flux for otherwise similar properties. The post- to pre-shock velocity depends on the density ratio between the jet and the target, but 
for high Mach numbers and a  polytropic index $\gamma$ of 5/3, $v_{pre}/v_{post}=4$ \citep{Draine_1993}. Therefore, high shock velocities always go together with high post-shock velocities, and the post-shock plasma will show some proper motion even if the obstacle is stationary w.r.t. the jet flow.

The mass-loss rate of the jet is 
\begin{equation}
\dot{M} = m_H\,v\,n\,\pi\,r^2
\end{equation}
where we have assumed a flow without voids (i.e., filling factor 1.0). Using Eq.~\ref{eq:L}, 
we can write
\begin{eqnarray}
\dot{M} & = &  m_H\,v \frac{L}{d_{cool}\, n\,\Lambda(T)}\\
        & \approx & 2\times10^{-10} \,M_\odot\,\textrm{yr}^{-1} \left( \frac{v_{flow}}{300\,\textrm{km}\,\textrm{s}^{-1}} \right) \left( \frac{300\,\textrm{km}\,\textrm{s}^{-1}}{v_{shock}} \right)^{4.5} \nonumber \\
        && \hspace*{3cm}\cdot \left(\frac{3\times10^{-23} \textrm{erg}\, \textrm{s}^{-1} \textrm{cm}^{-3}}{\Lambda(T)}\right)  \left( \frac{L}{10^{29}\,\textrm{erg}\,\textrm{s}^{-1}} \right)
\end{eqnarray}
which suggests that a small fraction of the total mass loss in protostellar jets is sufficient to power the X-rays. 
The above estimates are rough order of magnitude estimates as they ignore the three dimensional structure and evolution of the post-shock plasma. HD simulations suggest that these simple analytic expressions are correct within about an order of magnitude for homogeneous jets ramming into a homogeneous ambient medium \cite{Raga_2002}. 

Early simulations of the jet-powered X-ray emission concentrated on a single shock. To better capture the knotty jet structure, a number of more and more sophisticated (M)HD models\footnote{Most models intended to investigate the knotty structure of jets ignore the effect of the magnetic field as it is considered to be negligible at high distances from the star.} 
have been constructed to better understand the conditions in which fast shocks in jets produce X-rays. 
In these simulations, individual knots along the jet axis are caused by mutual interactions between clumps ejected at different epochs, i.e., they reflect some variability in the ejection process, similar to the optical jet.
To reproduce the X-ray emission, however, the simulations assumed random flow velocities ranging from 10 up to 5\,000\,km\,s$^{-1}$ and the X-ray emission is generated where the fastest material rams into slower material ejected earlier so that the post-shock plasma radiating the X-ray emission will have a significant proper motion of
between 300 and 3\,000\,km\,s$^{-1}$ \cite{Bonito_2010a,Bonito_2010b}.
These simulations also show that higher shock velocities and stronger X-ray emission are achieved for light jets, that is, jets with a lower density than the ambient medium \cite[see also][]{Bonito_2007}. This also applies to the simulations of a jet ramming into dense material at some distance to the driving source, where heavy jets require very high velocities to produce significant X-ray emission  \citep[$\gtrsim1\,000$\,km\,s$^{-1}$, e.g.,][]{Lopez_2015}. It is unclear if such high velocities are present in protostellar jets and we envision that a detailed comparison between the X-ray emission and the optical emission from near-simultaneous data will help to further constrain the physical conditions in the X-ray emitting plasma and, thus, the conditions which result in X-ray emission from protostellar jets.

\begin{figure}[t]

\includegraphics[width=\textwidth]{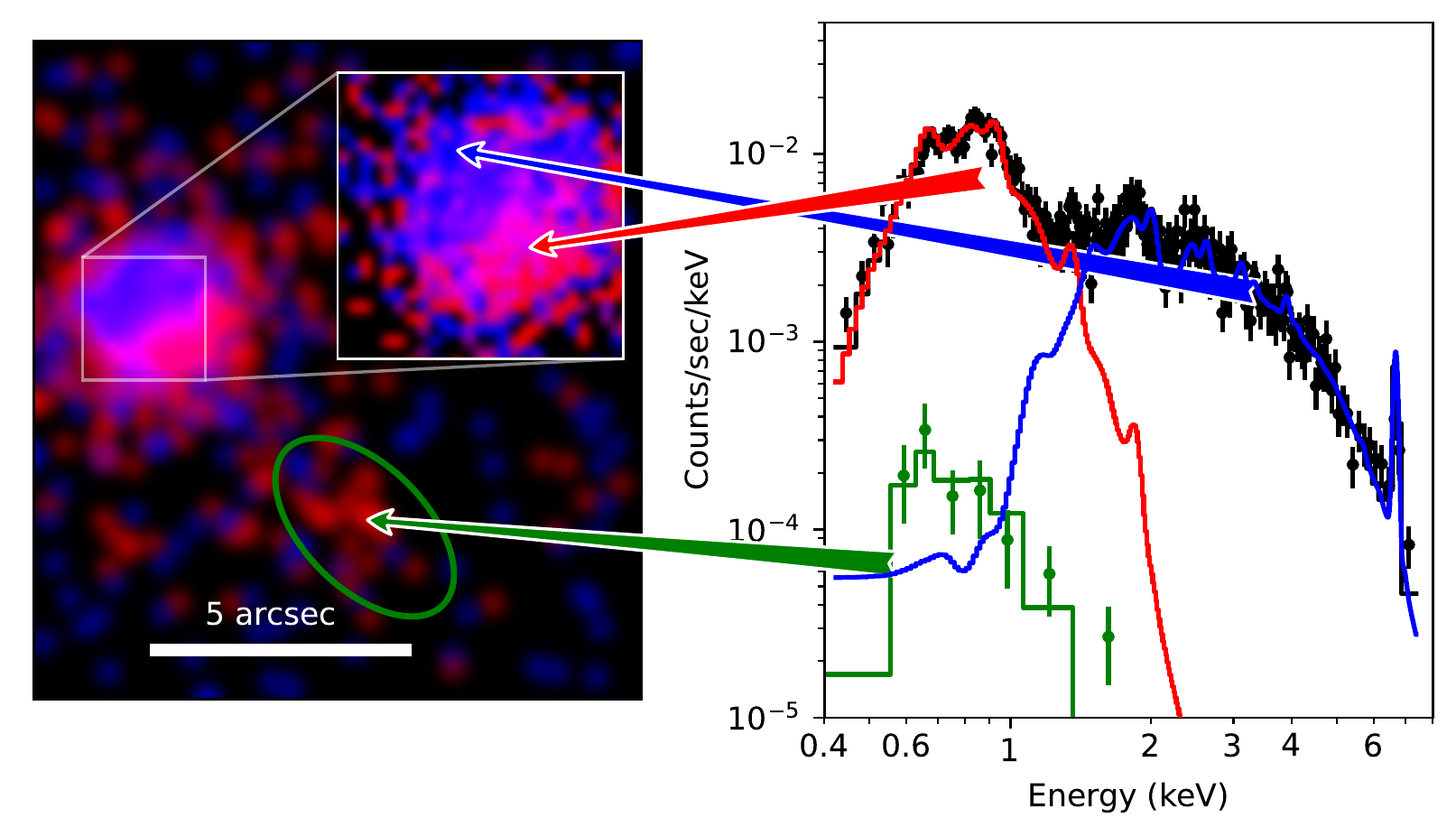}
\caption{{\bf Left: } \emph{Chandra} X-ray image of the DG~Tau system. Individual photons are shown in two bands (red: 0.4-1.2~keV, blue: 1.2-7.0~keV). Areas where both soft and hard photons are seen show up purple. There are three emission components: resolved jet (inside the green ellipse) and two components that seem to overlap because of \emph{Chandra}'s PSF: The star itself with hard, coronal emission (blue) and the jet base with soft X-ray emission (red) - see inset for a zoom-in. Both the main image and the inset are smoothed  and the color stretch is non-linear and chosen for visual clarity.
         {\bf Right: } Spectra from the central source (black) with two components (blue: a hard, highly absorbed, coronal source and a softer, less-absorbed source (red). Green is the faint emission of the resolved jet. Arrows connect spectral components with features in the image. \label{fig:dg_tau_X}}
\end{figure}

\subsubsection{X-rays from the jet base}
In addition to X-ray emission from jet knots at some distance to the driving sources, there have been a number of observations of X-ray emission from the jet base. The property that sets this X-ray emission from the jet close to the driving source ($d\lesssim100\,$au), called Jet Base X-ray emission (JeBaX) in the following, apart from the jets with X-ray emission from the outer jet knots is, however, the  evolution  and not the location of the X-ray emission. The JeBaX is stationary and no proper motion of individual X-ray emitting knots is seen, as is characteristic for optical and X-ray emitting knots in protostellar jets, despite multi-epoch X-ray observations of sufficient spatial resolution.

JeBaX examples are
HH~154 \citep{Favata_2002,Favata_2006,Schneider_2011,Bonito_2011}, DG Tau \citep[see Fig.~\ref{fig:dg_tau_X};][]{Guedel_2005,Guedel_2008,Schneider_2008}, HH 540 \citep{Kastner_2005}, and perhaps the inner component of the HD~163296 jet, because part of the cool plasma originates above the stellar surface although no spatial offset is detectable \cite{Guenther_2013}. The extended X-ray emission seen in RY Tau  may considered for this group, too,  although its faint jet X-ray emission is seen at more than 140\,au from the star \citep{Skinner_2011}. Also, the so-called two absorber X-ray (TAX) sources belong into this class as their X-ray spectra are composed of a strongly absorbed hard component associated with stellar coronal emission and a weakly absorbed soft component associated with the jet \citep[][]{Guedel_2007}; DG~Tau is also a TAX source with the soft X-ray emission $\lesssim1$\,keV subject to much less absorption than the hard component (see Fig.~\ref{fig:dg_tau_X} right). Flaring in TAX sources affects only the hard and not the soft component, which is further evidence for two different emitting regions being responsible for the soft and hard components \citep{Guedel_2007}; and against a scattering origin of the extended X-ray emission. 

In the following, we concentrate on DG~Tau and HH~154, because they have been observed several times in X-rays and
the observational situation for the other systems is inferior. We expect, however, that the physical processes leading to the jet's X-ray emission are very similar. In particular, there are no reasons to believe that the spatially unresolved TAX sources differ substantially from DG~Tau.
And although the evolutionary statuses of the different driving sources differ (e.g., L1551~IRS\,5 is often considered a class~I binary while DG~Tau is a class~II object), we assume in the following that the mechanisms generating the jet's X-ray emission are very similar in all these systems. We note, that JeBaX may have a relevant contribution to the ionization of the disk surface inducing disk accretion instabilities \cite{Balbus_1991} and altering the disk chemistry \cite{Glassgold_2004}.

Figure~\ref{fig:hh154} shows that the spatial distribution of the X-ray emission in HH~154 (also known as the jet from L1551~IRS\,5) is compatible with a source constant in flux and morphology  \cite{Schneider_2011,Bonito_2011}. The same applies to the  X-ray emission of the DG~Tau jet, where 
a careful analysis of the relative astrometry between the soft (jet) and hard (coronal) X-ray emission shows that the (deprojected) spatial offset amounts to about 40\,au in all available  \emph{Chandra}  epochs \citep[see Fig.~\ref{fig:dg_tau_X} and ][]{Schneider_2008}, which remains true in more recent observations \citep{Guedel_2011}. Therefore, JeBaX is observed over decades from the same location within the jet while the proper motion of individual X-ray emitting knots would have been easily detectable, i.e., the JeBaX is stationary.

The spatial extent of the X-ray emission differs between HH~154 and DG~Tau. In
HH~154, X-rays are seen out to about 420\,au with a peak at 140\,au \citep{Schneider_2011} while no significant source extension is seen for the DG~Tau soft comoponent limiting its size along the jet axis to $\lesssim100\,$au \citep{Schneider_2008}. For HH~154, \citet{Schneider_2011} spatially resolve the spectral properties of the X-ray emission and show that the plasma cools from about 8\,MK to 5\,MK within about 300\,au with a similar trend in all three epochs \cite[2001, 2005, 2009, see also][]{Bonito_2011}.

The X-ray luminosities are $\sim10^{29}$\,erg\,s$^{-1}$ for both jets with emission measures of $8\times10^{51}\,$cm$^{-3}$ (HH~154) and $3\times10^{52}$\,cm$^{-3}$ (DG Tau), and even though the inner X-ray emission region in DG~Tau is not resolved, we can derive a lower limit for the electron density using some reasonable assumptions for the spatial extent (100\,au in length and a radius of 35\,au) through 
\begin{equation}
n_e \gtrsim \sqrt{\frac{EM}{V}}\,,
\end{equation}
which results in $n_e>10^3\,$cm$^{-3}$  (HH 154) and $>3\times10^3\,$cm$^{-3}$ (DG Tau). Radiative cooling of a hot plasma with these densities is inefficient  (see Eq.~\ref{eq:dcool}) and any expansion, e.g. while moving away from the driving source, likely dominates the cooling and the X-ray flux reduction \citep{Guedel_2008, Schneider_2011}. 

To explain the JeBaX, one needs a mechanism that constantly (over the decade time scale)  heats the plasma close to the driving source to temperatures resulting in (detectable) X-ray emission ($T\gtrsim3\times10^6$\,K). This plasma then cools while traveling along the jet axis. Such a behavior differs strongly from expectations for an individual plasma blob heated by a strong shock as the resulting, hot post-shock plasma would show detectable proper motion. Furthermore, the shock heating would need to be very stable always heating the same ``amount'' of plasma to the same temperature. Such features are quite unique within the jet context and a number of different scenarios have been proposed to explain the stationary X-ray emission from protostellar jets.

\subsubsection{Origin of the X-ray emission at the jet base}

In analogy with the X-ray emission at larger distances, shocks have been proposed to cause the X-ray emission at the jet base, too. Due to the location of the JeBaX close to the driving source, shocks  are conceivable that  naturally explain the stationary nature of the X-ray emission. For example, an obstacle in the outflow path, which redirects the flow \citep[either the magnetic field or the dense ambient medium, see][]{Bally_2003,Guenther_2014}, or processes related to jet collimation \citep[e.g.][]{Schneider_2011}. 

A scenario proposed to describe the stationary X-ray emission of HH~154 is a diamond shock forming at the base of the jet \cite{Bonito_2011}; in this case, a shock that forms after a 1\,500\,km\,s$^{-1}$ fast jet with $n=300\,$cm$^{-3}$ exits a 200\,au wide fiducial nozzle potentially representing dense circumstellar material or a magnetic nozzle and rams into a dense ambient medium \citep[$n=3\times10^3\,$cm$^{-3}$,][]{Bonito_2011}.

\begin{figure}[t]
    \centering
    \includegraphics[width=12cm]{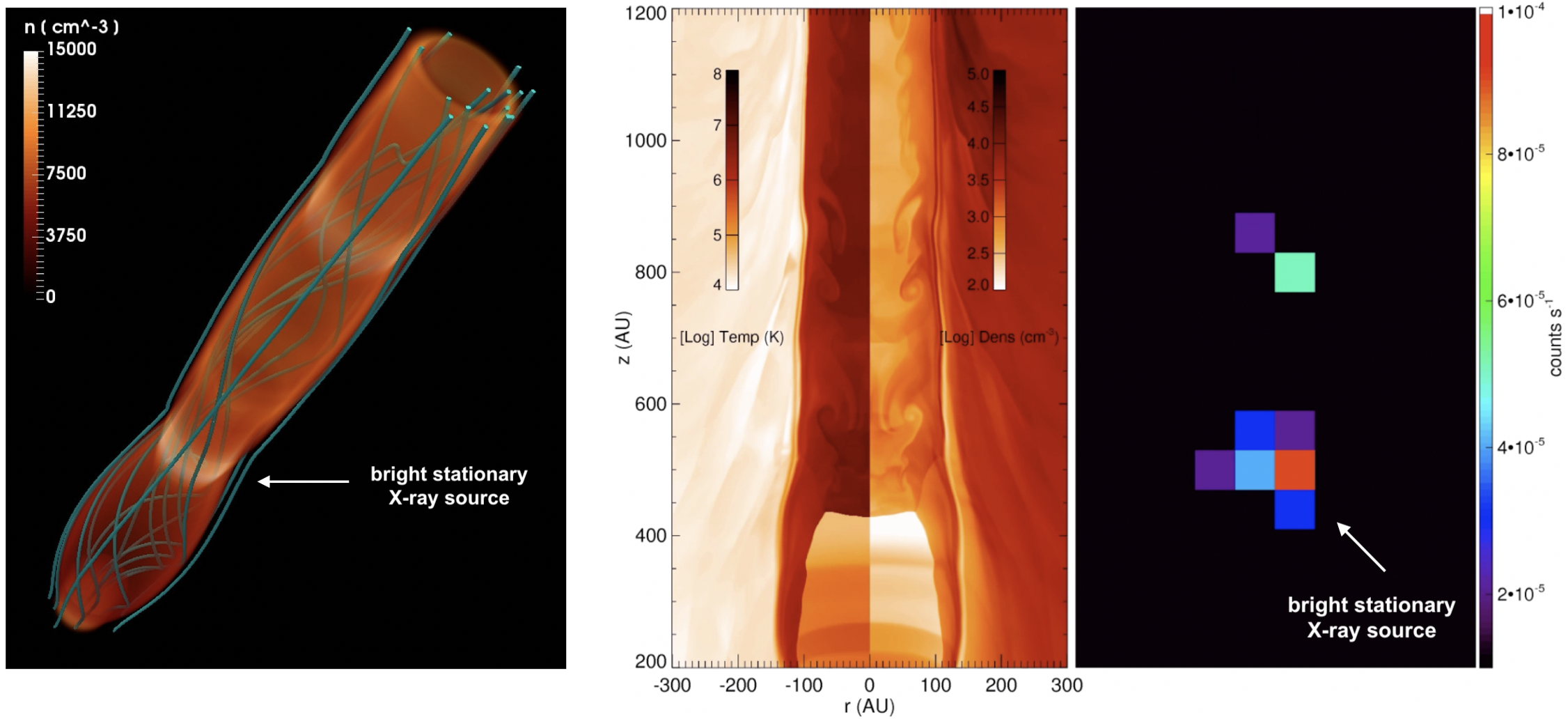}
    \caption{{\bf Left}: Three-dimensional representation of the density distribution and the magnetic field configuration (blue lines) 
    for the model M8 from \cite{Ustamujic_2016} after $\approx 50$~yr 
    of evolution. Figure credit: \citet{Ustamujic_2016}, A\&A, 596, 99, 2016, reproduced with permission © ESO.
    \textbf{Right}: Two-dimensional maps of temperature \emph{(left half-panel on 
    the left)}, density \emph{(right half-panel on the left)}, and synthetic 
    X-ray count rate \emph{(right panel)} in the [$0.3–4$]~keV band, with 
    macropixel resolution of $0.5\arcsec$ (\emph{Chandra}'s  native resolution),
    for the best fitting light jet model for HH~154. Figure credit: \citet{Ustamujic_2018}, A\&A, 615, 124, 2018, reproduced with permission © ESO.}
    \label{fig:ustamujic}
\end{figure}

Replacing the fiducial nozzle with a magnetic field, \citet{Ustamujic_2016} performed  2.5D MHD numerical simulations for a continuously driven jet ramming into a magnetized medium taking into account magnetic-field oriented thermal conduction  and radiative losses. The magnetic nozzle at the jet base leads to the formation of an X-ray emitting stationary shock diamond at its exit as well as to flow re-collimation \citep[see 3D reconstruction in left panel in Fig.~\ref{fig:ustamujic} and][]{Ustamujic_2016}. The shock reaches temperatures of few million degrees and densities of $\sim 10^4$cm$^{-3}$ (see left panel in Fig.~\ref{fig:ustamujic}) and is stationary over the time covered by the simulations ($\approx$40-60~yr). This configuration requires significantly lower flow velocities than previous models ($\sim$500\,km\,s$^{-1}$ vs $>1\,000\,$km\,s$^{-1}$)
to reproduce the observed X-ray emission, because flow compression by the ambient magnetic field contributes to the heating.

This shock diamond is even rather stable against perturbations in a pulsed jet scenario \cite{Ustamujic_2018} and the bright stationary  X-ray source is still present (see right panel in Fig.~\ref{fig:ustamujic}). Interestingly, similar collimation (magnetic) mechanisms form a quasi-stationary X-ray emitting shock at the base of the jet for light (appropriate for, e.g., HH~154) and heavy (such as, e.g., DG~Tau) jets \cite{Ustamujic_2018}. Most importantly, however, the synthetic observations derived from the simulations by \citet{Ustamujic_2018} are in good agreement with the actual X-ray data (see right panel in Fig.~\ref{fig:ustamujic}). The feasibility of the stationary-shock scenario at the base of the jet has also been demonstrated through sophisticated magnetized laser-plasma laboratory experiments \cite{Albertazzi_2014, PhysRevLett.119.255002} and is thus a promising scenario for the JeBaX.

In contrast to MHD simulations starting after the jet has been already accelerated,  \citet{Guenther_2014} studied semi-analytically how a fast inner outflow, a stellar wind, interacts with a surrounding, slower outflow launched from the disk. The stellar wind initially has a larger opening angle, but is collimated by the magnetic and thermodynamic pressure of the disk wind. A model assuming a stellar wind velocity of 840\,km\,s$^{-1}$, a launch radius of 0.1\,au, and a mass-loss rate of $5\times10^{-10}\,M_\odot$\,yr$^{-1}$ produces X-ray emission with spatial scales and luminosities compatible with observations \cite{Guenther_2014}.

All models relying only on shocks for heating the outflowing plasma to X-ray emitting temperatures require high flow and shock velocities, which are, however, not seen at other wavelengths. Therefore, the jet's magnetic field may contribute to the heating of the jet, either by facilitating re-collimation of the flow as for the latest diamond shock models, or through current dissipation and reconnection events \cite{Guedel_2007, Schneider_2013a}.
In particular, the disk atmosphere may be magnetically heated similar to the nanoflare model for the heating of the Sun's corona \cite{Takasao_2017}. Magnetic buoyancy leads to magnetic loops rising to large heights in the inner disk where energy is released in reconnection events. A disk magnetic field of 1\,G may be sufficient to produce the observed X-ray luminosity and temperature. In this model, the jet base is the X-ray brightest region and cooling by radiation and expansion may be sufficiently suppressed to produce X-rays at some distance to the launch region ($\gtrsim40$\,au). This suppression of (too) quick radiative or adiabatic cooling necessitate flow velocity in excess of 300\,km\,s$^{-1}$, because the heating takes place close to the disk in this model. Hence, the X-ray emission would trace the fast jet even without requiring shocks to provide the thermal energy in this model \cite{Takasao_2017}. 

\subsubsection{Comparison with other jet tracers}
How does the X-ray emission compare with data at other wavelengths? First, 
evidence for stationary structures also exists in tracers of lower temperature plasma, e.g., [Fe~{\sc ii}]~1.644\,$\mu$m  \citep{White_2014} and [O~{\sc i}]~$6300\,$\AA{} data \citep[][]{Schneider_2013a}, neither of which is located clearly downstream of the X-ray emission as expected if they represent material traveling along the jet that cooled after being so hot that it radiates X-rays. 

Second, optical and NIR jet observations suggest shock velocities of only 30--100\,km\,s$^{-1}$ close to the source \citep[e.g., ][]{Lavalley_2000, Hartigan_2007}. The detection of [O~{\sc iii}]~$5007\,$\AA{} emission in some shocks may imply temperatures of $\log~T\sim5$~K although the bulk emission still appears to result from lower temperature plasma \citep[e.g.,][]{Bacciotti_2011,Nisini_2016}. On the other hand, the shock velocity appears to increase with post-shock velocity and there is evidence for high-velocity ($\sim400\,$km\,s$^{-1}$) material close to the star in some sources from absorption studies, e.g., sub-continuum absorption in He~{\sc i}\,$10830\,$\AA{} \citep[e.g.][]{Edwards_2006} or FUV lines like C~{\sc ii} \citep{Xu_2021}. However, typical velocities are slower and even flow velocities of 400\,km\,s$^{-1}$ would still require a stationary obstacle for shock velocities sufficiently fast for X-ray emission. Therefore, the extremely fast flows needed to produce the JeBaX exclusively via shocks appear to have escaped direct detection so far. 

Third, observations of emission tracing the temperature range intermediate between the ``classical''  and the X-ray emitting jet have been obtained with HST (C~{\sc iv}, [Ne~{\sc iii}]). Long-slit HST C~{\sc iv} data of the DG~Tau and RY~Tau  jets show that C~{\sc iv} emission is located close to the star \cite{Schneider_2013a, Skinner_2018}; the C~{\sc iv} emission in the DG Tau jet is concentrated in a discrete structure overlapping with the X-ray centroid ($\approx0.2$\,arcsec)  while 
the C~{\sc iv} flux monotonically decreases away from the source in RY~Tau. In both jets, the deprojected C~{\sc iv} velocities decrease from about 300\,km\,s$^{-1}$ close to the star to roughly 200\,km\,s$^{-1}$, which is somewhat faster than the optical emission in both jets but only by a small factor \cite[$\approx 1.5$, see ][]{Schneider_2013a, Skinner_2018}.

Highly ionized plasma is also traced by [Ne~{\sc iii}]~$3869\,$\AA{} emission\footnote{A shock velocity of $\gtrsim100\,$km\,s$^{-1}$ is required to produce [Ne~{\sc iii}] in shocks.} in the Sz~102 jet \citep{Liu_2021}. The [Ne~{\sc iii}] emission is spatially extended along the jet axis, and smoothly decreases in flux away from the star, like the C~{\sc iv} emission in RY~Tau. Therefore, \citet{Liu_2021} prefer a scenario in which the high ionization is produced very close to the star, likely due to X-ray irradiation,  and remains frozen in the jet as the recombination time scale appears compatible with the spatial extend of the [Ne~{\sc iii}] emission and the measured flow velocity of 250--300\,km\,s$^{-1}$.

Also, mass-loss rates estimated for the $10^5$\,K jets \citep[$\sim10^{-9}\,M_\odot$\,yr$^{-1}$,][]{Schneider_2013a} may be regarded as intermediate between the ``classical'' \citep[$\sim10^{-8}\,M_\odot$\,yr$^{-1}$, e.g., ][]{Bacciotti_2000} and X-ray emitting jets ($\dot{M}_{out}(\textrm{X-ray})\sim10^{-10}\,M_\odot$\,yr$^{-1}$, see above). 

\subsection{Towards a coherent model for X-ray emission from protostellar jets}
Protostellar outflows carry sufficient kinetic energy to power the X-ray emission via shocks, and different shock scenarios have been explored. 
The X-ray emission from jet knots has all the properties expected from shock-heated plasma; it shows proper motion, is spatially associated with lower temperature plasma, and often appears where the flow is fastest or where other high excitation tracers such as [O~{\sc iii}] are seen. And although relatively high flow and shock velocities are needed to cause X-ray emission from jet knots, no fundamental challenges for the shock model have been identified so far.

Explaining the JeBaX is more challenging, because it differs considerably from the optical jet. The location close to the driving source, likely in the jet acceleration and collimation region, appears to be crucial to explain the  stationary nature of the X-ray emission. A viable scenario for the JeBaX may be based on the fastest jet component, which carries only a fraction of the mass-loss and, thus, may have escaped direct detection at other wavelengths so far. Extending the trend of increasing shock velocity with increasing flow velocity seen at (much) lower plasma temperatures may be compatible with shock-powered X-ray emission provided that (magnetic) re-collimation contributes to the heating. 

In this picture, the X-ray and C~{\sc iv} emission represent the innermost and fastest components of the onion-like structure of protostellar jets extending the regularly detected low-, medium-, and high-velocity components to even higher velocities and temperatures. 

Producing the observed X-ray emission with  one or a number of discrete shocks without any magnetic effects appears challenging, because measured flow velocities are on the low end of the shock velocity range required to produce the X-ray emission. And although this issue can be partly  circumvented by assuming that (a) velocities are post-shock velocities with much higher initial flow velocities and (b) that only a fraction of the flow participates in the X-ray production, we think that incorporating magnetic effects provides a much more natural explanation for the stationary nature of the JeBaX emission, because this directly relates its spatial location to the magnetic outflow (re-) collimation. 

Lastly, we note that the stark contrast between the JeBaX and other outflow tracers remains challenging, but we also note that indications for stationary emission close to the driving source is not exclusively seen in the X-ray regime. Still, other mechanisms like magnetic heating may be required to explain the high temperatures at velocities only slightly larger than the ``classical'' jet where shock velocities fall short of the ones required for the X-ray emission by almost an order of magnitude. 

Additional observations and simulations are needed to investigate the interplay between the different temperature components in protostellar jets; this approach may also constrain the magnetic field in the jet launching region, which is linked to the  magnetic field in the protoplanetary disk where planets form.

\section{Summary \& Outlook}

X-ray data have shown new, surprising results for the dynamics of material close to PMS stars, i.e., for accretion and outflow processes. Accretion produces clear observational signatures in the X-ray spectra of young stellar systems, namely cool emission from a high density plasma as shown by \emph{XMM-Newton} and \emph{Chandra} grating spectra of the nearest and X-ray brightest systems. Similarly, \emph{XMM-Newton} and \emph{Chandra} observations have revealed about one dozen systems with  X-rays from outflows, likely from the systems with the fastest and most powerful jets.
Both processes are strong sources of ionizing radiation and, thus, may impact the entire star formation process including how planets form.

The existing observations, however, suffer from low photon numbers and modest spectral resolution. Day-long integrations are needed to obtain a density estimate for the accretion-generated plasma in the nearest, brightest systems, or to get a temperature estimate for the jet plasma emitting X-rays. We expect that the comparison of the X-ray data with tracers for lower temperature plasma at other wavelengths offers great potential, but the rather long observations currently required for the X-ray spectra pose issues for the intrinsically variable accretion process, because the accompanying, complementary data often provide only snapshots of the accretion state during the X-ray observations. Also, X-ray emitting jets are just too faint to measure line shifts, which would inform us about the process that generates the X-rays in the first place. 

Modern numerical models for both the accretion and the outflow process simulate the flows in more than one dimension, study the influence of the magnetic field and add methods to tread optical depth and non-ideal MHD effects. Simulations of the accretion column consistently produce a type of outer shell of material that escapes the accretion funnel and obscures the infalling funnel. Jet simulations have progressed to provide scenarios compatible with measured outflow speeds and existing X-ray data but launching conditions are not well constrained yet. In the future, improved observations in the X-rays and in other bands, will provide more stringent comparisons for the models and thus hopefully elucidate the launching and collimation mechanism for protostellar jets.

New X-ray facilities like the ESA mission \emph{Athena+} will provide the sensitivity to not only enlarge the samples, but will provide those diagnostics that are needed to test models. Measuring the line shifts for the accretion- and jet-powered X-rays will be among the most important tools to decipher the underlying physics as measuring velocities for the X-ray emitting plasma can directly inform us about the physical origin of the plasma. Even more can be learnt from spectrally resolving the accretion and outflow emission lines. And while \emph{Athena+} will be instrumental for pushing our understanding of the hot and energetic processes of young stellar systems, X-ray observatories providing higher resolution spectra have been proposed, too, which are more appropriate for measuring line shapes and not just shifts, like \emph{Arcus} or \emph{Lynx} proposed for US missions.

Accretion and outflows are defining characterics of star and planet formation, and both processes genuinely generate X-rays. These X-rays control many of the traditionally studied tracers, e.g., through the heating of the infalling  material or by affecting disk ionization. Any model of star and planet formation must, therefore, also include the most energetic part, the X-ray emission.


\section{Cross-References}
\begin{itemize}
    \item Stellar Coronae  - J.Drake and B. Stelzer  
    \item Star forming regions - S.Sciortino   
    \item Young Moving Groups - J.Kastner and D. Principe
\end{itemize}

\bibliographystyle{aa_mod}
\bibliography{bib.bib}

\end{document}